\documentclass{ws-ijmpa}
\usepackage{soul, color}
\usepackage{graphicx, subfigure, multirow, mcite, amsmath, amstext, collref, hyperref}
\usepackage[super,compress]{cite}
\hypersetup{colorlinks,urlcolor=black,citecolor=black,linkcolor=black,filecolor=black}
\usepackage{breakurl}
\graphicspath{{figures/}}

\begin{document}

\title{\bf Universal mass limits on gluino and third-generation squarks in the context of Natural-like SUSY spectra \\
\vspace{0.5em}}

\author{{\bf O.~Buchmueller}, {\bf J.~Marrouche}}
\address{High\,Energy\,Physics\,Group,\,Blackett\,Laboratory\,Imperial\,College,\\Prince\,Consort\,Road,\,London,\,SW7\,2AZ,\,UK}

\maketitle

\begin{abstract}
In this paper, we present results based on a combination of four inclusive topology searches for supersymmetry (SUSY) from the CMS experiment, and use this to determine universal mass limits on gluino and third-generation squarks in the context of Natural-like SUSY spectra. This class of sparticle spectra is inspired by the argument of naturalness, which originates from both consideration of fine-tuning arguments, and the need to satisfy current experimental search constraints. The class of Natural-like SUSY spectra considered follows the typical Natural SUSY model made up of a gluino, third-generation squarks, and higgsino sparticles, but is extended to more complex spectra containing sleptons. We show that the limits obtained from the combination of inclusive topology searches are far more stable than those from individual searches, with respect to the assumed underlying complexity of the spectra, and hence these limits can be considered as universal mass limits on gluino and third-generation squarks, defined in the context of this broad class of Natural-like SUSY spectra. Furthermore, we present our results using a simple colour scheme that allows a straightforward interpretation of any Natural-like SUSY spectrum with our limits.

Complementing the final results of the 2011 searches based on 5~$\textrm{fb}^{-1}$ of integrated luminosity, with the first published results from the 2012 searches using approximately 11~$\textrm{fb}^{-1}$, we find that gluinos with a mass of $m_{\tilde{g}}\approx$~1050~GeV and third-generation top and bottom squarks with masses of $m_{\tilde{3G}}\approx$~575~GeV are excluded for low masses of the lightest SUSY particle (LSP). These limits weaken to $m_{\tilde{g}}\approx$~600~GeV and $m_{\tilde{3G}}\approx$~450~GeV, when the mass of the LSP is increased to several hundred GeV. Based on this result, we establish a prediction of how these limits might evolve when the full 2012 data set is analysed, with both the CMS and ATLAS experiments combined. This outlook suggests that for low masses of the LSP, gluinos with a mass of $m_{\tilde{g}}\approx$~1150~GeV and third-generation squarks with masses of $m_{\tilde{3G}}\approx$~675~GeV are likely to be excluded. For high LSP masses, these limits are expected to decrease to $m_{\tilde{g}}\approx$~650~GeV and $m_{\tilde{3G}}\approx$~500~GeV.
Therefore, despite the fact that the LHC already probes a significant region of the SUSY parameter space, Natural SUSY scenarios based on rather stringent fine-tuning requirements may not be fully excluded by the data taken so far. This suggests that additional data is needed, to be recorded during the higher energy running of the LHC expected in 2015. The importance of combining relevant inclusive topology searches, in order to make the most universal interpretations possible, is a general recommendation for future experimental searches at the LHC.
\end{abstract}

\section{Introduction}

The landscape of supersymmetry (SUSY) searches has changed significantly since the Large Hadron Collider (LHC) at CERN began physics operation in 2010. Up until the end of 2011, the two general purpose experiments CMS and ATLAS had each collected approximately $5~\textrm{fb}^{-1}$ of integrated luminosity at a centre of mass energy of 7~TeV, and in 2012 a further $20~\textrm{fb}^{-1}$ was collected per experiment, but at a centre of mass energy of 8~TeV. While the full analysis of the 2012 data set is currently ongoing, both experiments have completed their SUSY searches for 2011. So far, no significant evidence for physics beyond the Standard Model (SM) has been found. Instead, the impressive number of SUSY searches have produced increasingly stronger limits on coloured super-particles (sparticles) decaying to missing energy~\cite{Khachatryan:2011tk,Chatrchyan:2012wa, Chatrchyan:2012sa, Chatrchyan:2012te, Chatrchyan:2012ola, CMS:2012un, Chatrchyan:2012pka, Chatrchyan:2012rg, Chatrchyan:2012jx, CMS:2012mfa, Chatrchyan:2012qka, Chatrchyan:2012mea, springerlink:10.1007/JHEP07(2012)167, springerlink:10.1140/epjc/s10052-012-2174-z, PhysRevD.87.012008, PhysRevLett.109.211802, PhysRevLett.109.211803, Aad2013879, Aad2013841, PhysRevD.86.092002, anotherATLASpaper, Aad201313, yetanotherATLASpaper, PhysRevLett.107.221804}, placing tight constraints on the available SUSY parameter space. 

However, the recent discovery of a Higgs-like resonance at the LHC~\cite{ATLASHIGGS, CMSHIGGS} is favoured by SUSY, in that its mass value of around 125 GeV falls within the narrow window allowed in the framework of the Minimal Supersymmetric Standard Model (MSSM), of between 115 and 135 GeV~\cite{springerlink:10.1140/epjc/s2003-01152-2, Carena200363}. This discovery, together with the observation of no direct experimental evidence for SUSY so far, has resulted in a shift of SUSY benchmark scenarios away from constrained SUSY models like the CMSSM~\cite{PhysRevLett.49.970, PhysRevLett.69.725, PhysRevD.49.6173}. The direction taken by both the CMS and ATLAS collaborations since then, has been to benchmark and optimise searches mainly in the context of Natural SUSY models~\cite{Barbieri198863, Romanino2000165, PhysRevLett.84.2322, Kitano200558, PhysRevD.73.095004, Giudice200619, 1126-6708-2009-10-061, Horton2010221}. These models avoid the direct search constraints from the LHC by assuming that the first and second generation squarks are very heavy, while at the same time postulating that the third-generation squarks are well below 1 TeV, in order to maintain the ability to solve the hierarchy problem~\cite{Weinberg:1975gm, Gildener:1976ai, Susskind:1978ms, Weinberg:1979bn} without significant fine-tuning. For this reason, the CMS and ATLAS experiments are now also providing interpretations in so-called simplified model spectra (SMS) scenarios~\cite{PhysRevD.79.075020,PhysRevD.79.015005}. 

Several phenomenological interpretations of recent experimental searches in the context of Natural SUSY models have already been performed~\cite{springerlink:10.1007/JHEP09(2012)035, springerlink:10.1007/JHEP03(2012)103, springerlink:10.1007/JHEP01(2012)074, springerlink:10.1007/JHEP02(2012)115, Baer:2012uy}.
In this paper, we interpret the latest experimental results in the context of Natural SUSY models by generating signal events using PYTHIA~\cite{1126-6708-2006-05-026}, and passing them through the DELPHES~\cite{Ovyn:2009tx} detector simulation package, with an appropriate data card emulating the response of the detector. Within our framework, we then implement the constraints of the relevant experimental searches to obtain yields in the respective signal regions. These signal yields are then confronted with the SM background yields and observations in data, as reported by the experimental searches, to set the appropriate limits on the gluino and third-generation squark masses. In order to have a consistent set of inclusive topology searches, and because of the familiarity of the authors with the CMS experiment, we confine our work to a representative set of CMS analyses spanning multiple final-state signatures. As we will see, for individual topology searches, the mass limits strongly depend on the complexity of the assumed underlying sparticle spectrum, thus preventing a model independent conclusion. However, when combining the most important topology searches, it is possible to establish limits that are almost independent of both the exact sparticle content and the allowed decay chains of the Natural SUSY model. We find that after combination of the search results, gluino and third-generation squark mass limits only depend on the assumption made about the mass of the lightest SUSY particle (LSP). Therefore, these limits can be expressed as a function of the LSP mass, exhibiting a similar dependence to limits defined in the context of SMS scenarios. However in this case, the limits are representative of a large class of Natural-like SUSY models, instead of only one simple spectrum and corresponding decay chain.
Furthermore, preliminary studies also indicate that extending our analysis to include the average first and second generation squark mass will enable universal limits on coloured sparticle production for an even broader class of SUSY spectra to be established. We are planning to follow-up on this extension in a future publication.
We would like to point out that the same program of work would be possible both with a study based on an alternative SUSY model set, such as the pMSSM~\cite{Cahill-Rowley:2013yla, PhysRevD.88.035002, Boehm:2013qva, epjC722169, PhysRevD.87.115012, PhysRevD.86.075025, Sekmen:2011cz}, and also by combining a consistent set of relevant ATLAS searches. In fact in previous work, searches from ATLAS have also been reinterpreted in a similar fashion within this same framework~\cite{Buchmueller:2012hv}. The overall aim of this paper is not to determine the most stringent universal limit, but to highlight the concept of how the universality of experimental limits may be achieved, and how the corresponding interpretations may be utilised in a wide-variety of SUSY model sets.

\section{Combining Inclusive SUSY Topology Searches}
\label{sec:search}
Assuming that R-parity is conserved~\cite{Farrar:1978xj}, SUSY particles such as squarks and gluinos are produced in pairs at the LHC and decay to the LSP, which is generally assumed to be a weakly interacting massive particle. This results in a final state signature which is both rich in jets and contains a significant amount of missing transverse energy. Depending on the details of the production mechanisms and decay chains involved, the general signature of jets and missing transverse energy can be accompanied by leptons and/or photons in the final state. For the studies in this paper, we assume that the LSP is the lightest neutralino, $\tilde{\chi}^{1}_{0}$.

For the first data taking campaign of the LHC in 2010, both the CMS and ATLAS experiments developed inclusive topology SUSY searches, which were characterised per event by the amount of missing transverse energy, the number of jets and hadronic activity, and the number of leptons as well as photons in the final state. By also ensuring a similar performance across all searches, this categorisation enabled the search strategy to be as model independent as possible, since the SUSY production cross sections at the LHC mainly depend on the masses of the particles considered (to a good approximation). At the same time, this also provided good sensitivity to a large variety of different SUSY production and decay topologies. For illustration, the 2010 CMS SUSY search categories are summarised in Table~\ref{tab:CMSsusysearches}.

\begin{table*}[!tbh!]
\tbl{An overview of the CMS SUSY search strategy as defined in 2010.}{
\begin{tabular}{| p{1.7cm} | p{1.7cm} | p{1.7cm} | p{1.7cm} | p{1.7cm} | p{1.7cm} | p{1.7cm} |}
\hline
0-lepton & 1-lepton & 2-leptons & 2-leptons & Multi-leptons & 2-photons & photon + lepton \\
\hline
Jets + MET & Single Lepton + Jets + MET & Opposite sign di-lepton pair + Jets + MET & Same sign di-lepton pair + Jets + MET & Multi Lepton & Di-Photon + Jets + MET & Photon + Lepton + MET \\
\hline
\end{tabular} \label{tab:CMSsusysearches} }
\end{table*}


The aspect of combining different search topologies to not only obtain the best but also the least model dependent limits on sparticle masses was less relevant in the first two years of LHC data taking. This was because limits were placed in constrained models like the CMSSM. In this class of model, the squarks are nearly degenerate in mass and therefore the exclusion of parameter space is mainly driven by gluino production, and the first (and to a lesser extent second) generation squark production. These processes possess the largest production cross sections in proton-proton collisions, and therefore the zero-lepton (and to a lesser extent the one-lepton) inclusive search, dominate the exclusion of parameter space, as they possess the highest sensitivity to these signatures. For this reason, a combination of all topology searches would not result in a vast improvement on the current limits and exclusion of parameter space, although the most stringent limits could only be obtained in this way.

However, while the combination of the inclusive topology searches would not be vastly beneficial for SUSY models with almost degenerate squark masses like the CMSSM, it is essential for Natural-like SUSY models in which the first and second generation squarks (i.e. the driver of the limits in the CMSSM) can be much heavier. This means that the limits on the SUSY parameter space relevant for Natural SUSY models (i.e. gluino and third-generation squark masses) depend more strongly on the details of the underlying sparticle spectrum. Thus, a priori, it is not clear which inclusive topology search, or set of inclusive topology searches, would be most relevant for SUSY models in which the first and second generation squarks are decoupled from the third-generation. As we will show in this paper, it is critical to combine all relevant inclusive topology searches in order to obtain limits on sparticle masses in Natural-like SUSY models which are not only the most sensitive, but also the least model dependent.



\section{Reinterpretation of inclusive SUSY searches}\label{sec:validation}
In order to perform an analysis of the Natural-like SUSY models to be considered for this paper, we have developed a framework based around the DELPHES detector simulation, which enables us to predict the signal expectations from multiple experimental searches, and interpret these expectations when combined with the SM background estimates and the LHC data as reported by the individual experimental searches. 
We do this by first generating events for a particular SUSY particle spectrum using the PYTHIA event generator. These events are then processed using the DELPHES fast detector simulation with an appropriate card to emulate the response of the CMS detector. We use NLO cross-sections obtained using PROSPINO~\cite{Beenakker:1996ed} throughout this paper. We then refer to journal publications in order to replicate the analyses and hence estimate the signal yields for each signal region of the following CMS searches: 0-lepton $\alpha_{T}$~\cite{Chatrchyan:2012wa}, 1-lepton $L_{p}$~\cite{Chatrchyan:2012ola}, opposite-sign dilepton~\cite{Chatrchyan:2012te}, and same-sign dilepton~\cite{Chatrchyan:2012sa}. All these searches are published using the entire 2011 dataset at $\sqrt{s}=7$~TeV. We make no attempt to reproduce the SM background estimates as these are normally data driven and rely on a deep understanding of the detector. Finally, we combine the signal yields, background estimates, and data observation and test the overall significance using the $CL_{s}$ test statistic~\cite{cls-pdg}.

\begin{figure*}[htb!]
\centering
\subfigure[]{\resizebox{7cm}{!}{\includegraphics{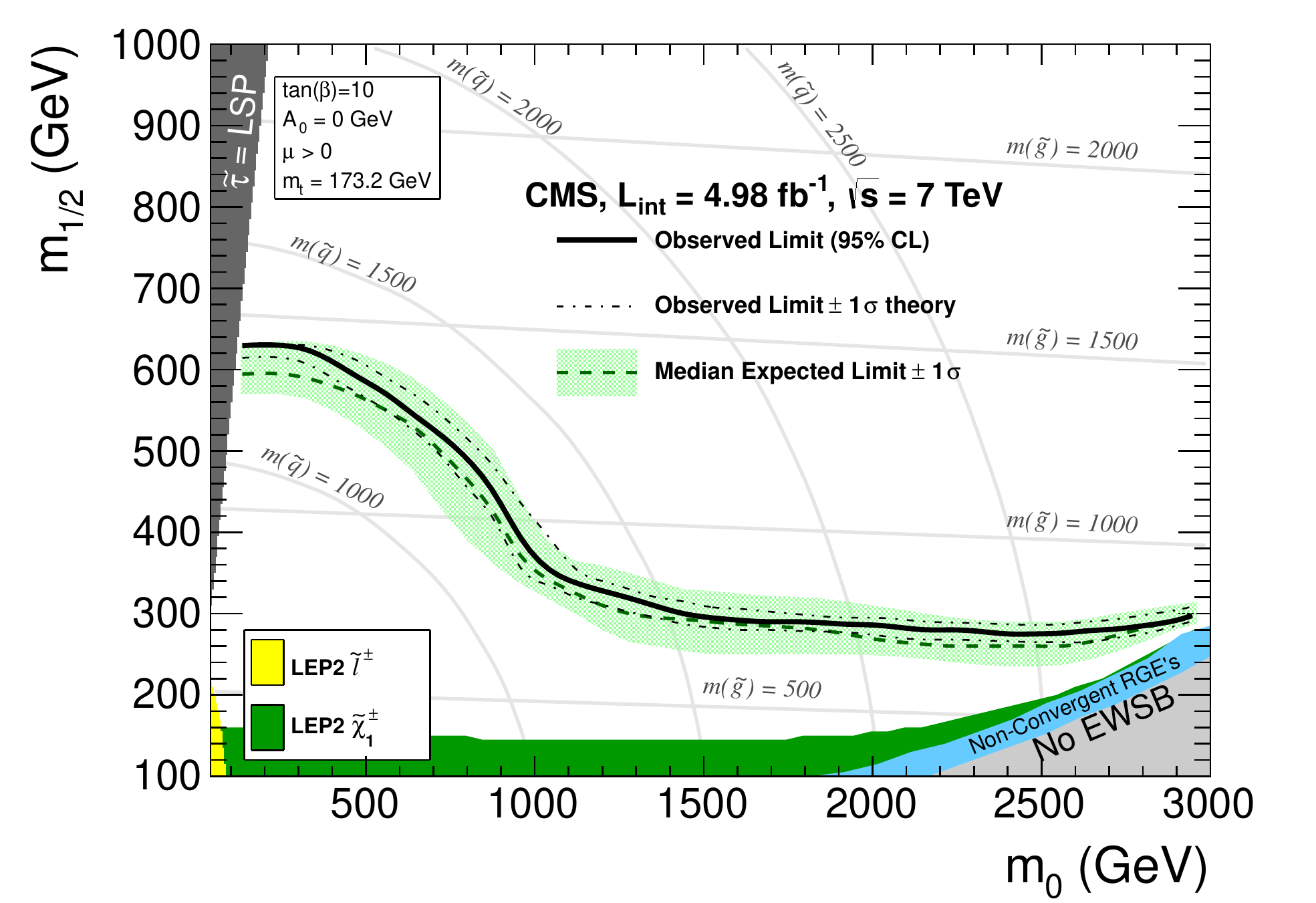}}\label{subfig:alphatlimit}}
\subfigure[]{\resizebox{5cm}{!}{\includegraphics{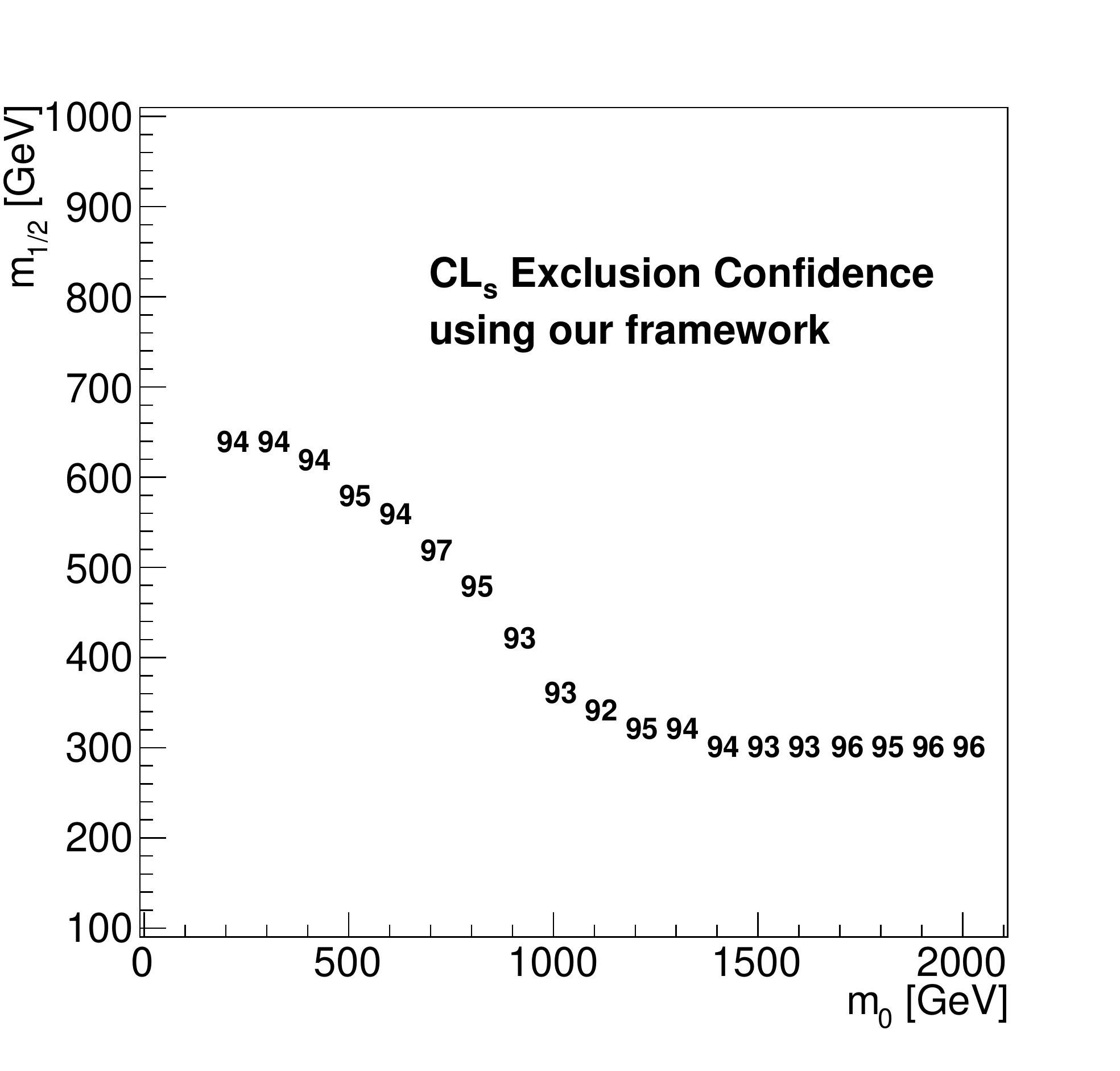}}\label{subfig:ouralphatlimit}}
\caption{\it The 95\% $CL_{s}$ exclusion confidence limit line as reported by CMS for the $\alpha_{T}$ search (left), and the corresponding $CL_{s}$ confidence level calculations using our framework for a set of test points along this same line (right). The numbers shown in Figure~\ref{subfig:ouralphatlimit} are given in percent.}
\label{fig:validationexample}
\end{figure*}

In order to validate the DELPHES detector simulation for our purposes, and also the implementation of the searches within our framework, we resort to trying to reproduce the reported CMSSM limits in the ($m_{0}$, $m_{1/2}$) plane for each analysis. We do this by scanning along the reported 95\% confidence level limit curves for each analysis in the ($m_{0}$, $m_{1/2}$) plane, using our framework to calculate the corresponding $CL_{s}$ exclusion confidence. Comparing this to the reported exclusion limits provides a stringent validation of our machinery in different parts of the CMSSM parameter space, where different production and decay chains are of relevance. For example, at low $m_0$, the limits are mainly driven by squark-squark production, typically yielding dijet final states, while at medium values of $m_0$, squark-gluino production dominates with 3-4 jets. At high values of $m_0$, the limits depend mainly on the gluino-gluino production mechanism, leading to jet multiplicities above four. Therefore, scanning in this plane is a good overall test for how effective our framework can reproduce the experimental searches for very different kinematic topologies of transverse missing energy signatures with different final states. An example of this scan in the ($m_0$, $m_{1/2}$) plane for the $\alpha_{T}$ inclusive search is shown in Figure~\ref{fig:validationexample}. As can be seen, each of our chosen test points closely reproduce the 95\% confidence level exclusion reported for this search.

\begin{figure*}[htb!]
\centering
\subfigure[]{\resizebox{6cm}{!}{\includegraphics{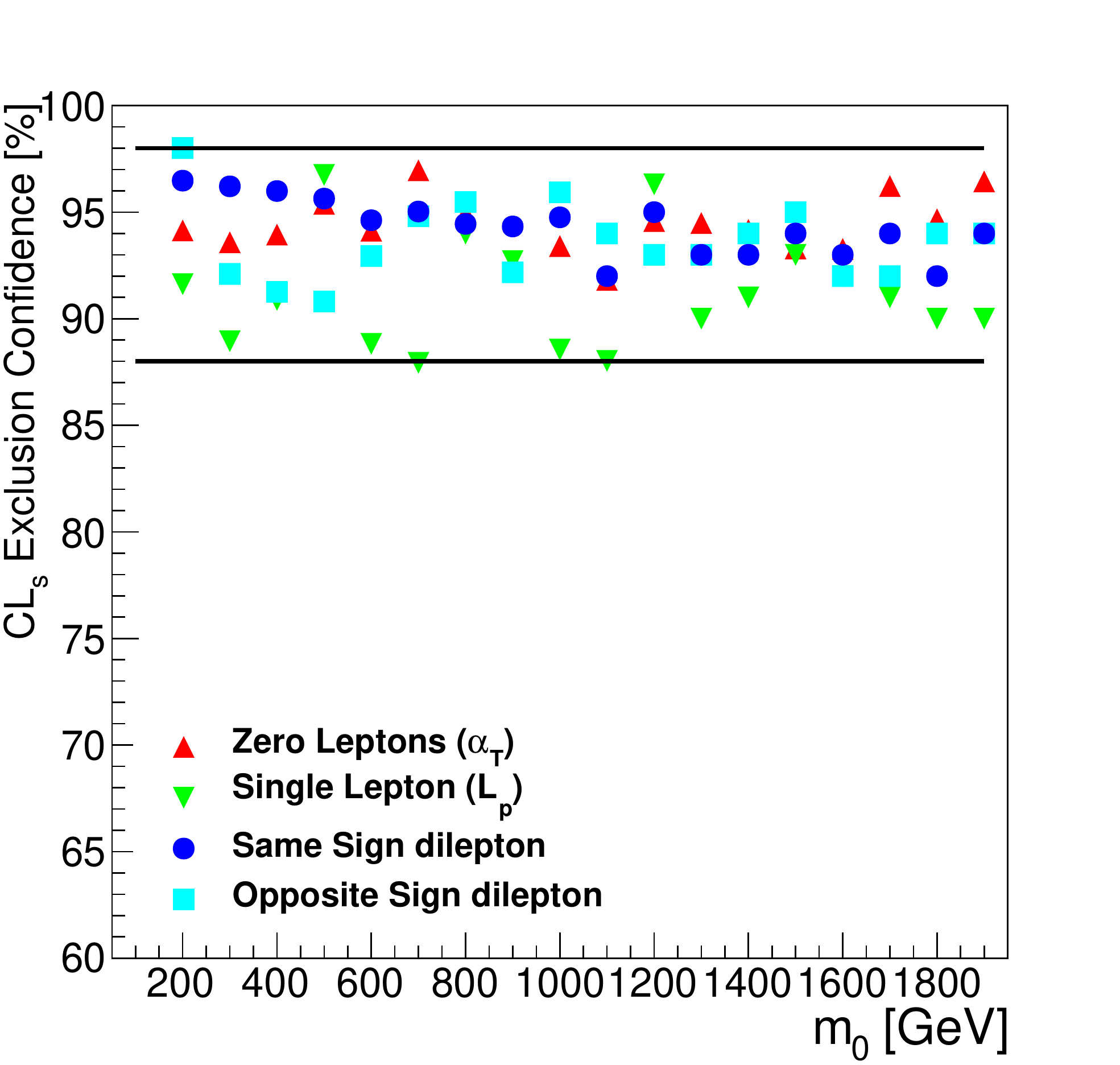}}\label{subfig:CLs_vs_m0}}
\subfigure[]{\resizebox{6cm}{!}{\includegraphics{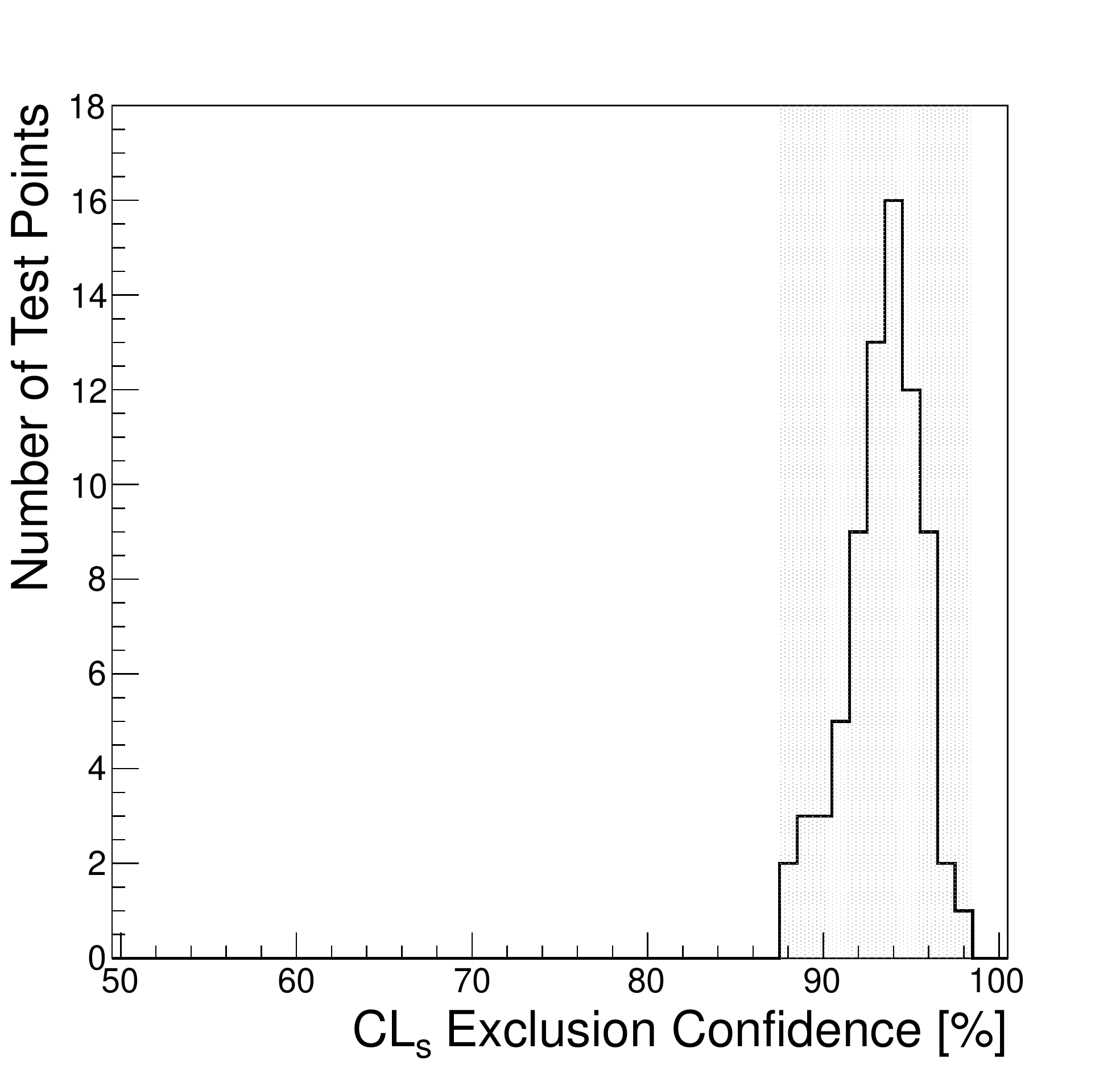}}\label{subfig:CLs}}
\caption{\it The $CL_{s}$ exclusion confidence levels as obtained using our framework, for CMSSM test points chosen along the reported 95\% $CL_{s}$ exclusion confidence level lines for each of the four CMS searches in their corresponding ($m_{0}$, $m_{1/2}$) planes. Figure~\ref{subfig:CLs_vs_m0} shows the result as a function of $m_0$, while Figure~\ref{subfig:CLs} displays the same information projected onto the exclusion confidence level axis. The two lines in Figure~\ref{subfig:CLs_vs_m0} and the grey band in Figure~\ref{subfig:CLs} represent the conservative range in which we are able to reproduce the experimental exclusion limits. This translates into an overall confidence level estimate of 95\%$^{+3\%}_{-7\%}$.}\label{fig:validation}
\end{figure*}

The complete results of this validation exercise for all searches are shown in Figure~\ref{fig:validation}. Figure~\ref{subfig:CLs_vs_m0} shows the calculated confidence level using our framework for each of the test points considered as a function of $m_{0}$, while Figure~\ref{subfig:CLs} shows these same test points projected onto the confidence level axis.

No significant systematic biases are observed, and the overall level of agreement achieved is used to determine a conservative estimate on the upper and lower bound on the definition of the 95\% confidence level in our analysis. This is shown by the two solid lines in Figure~\ref{subfig:CLs_vs_m0}, and the grey band in Figure~\ref{subfig:CLs}. The variation ranges maximally from a confidence level of 88\% to 98\%, but is mostly well contained within a few percent around the expected confidence level of 95\%. In the following, we assume that we are able to reproduce the experimental exclusion limits within 95\%$^{+3\%}_{-7\%}$. These uncertainties are later translated into uncertainties on extracted mass limits.

\begin{figure*}[htb!]
\centering
\resizebox{8cm}{!}{\includegraphics{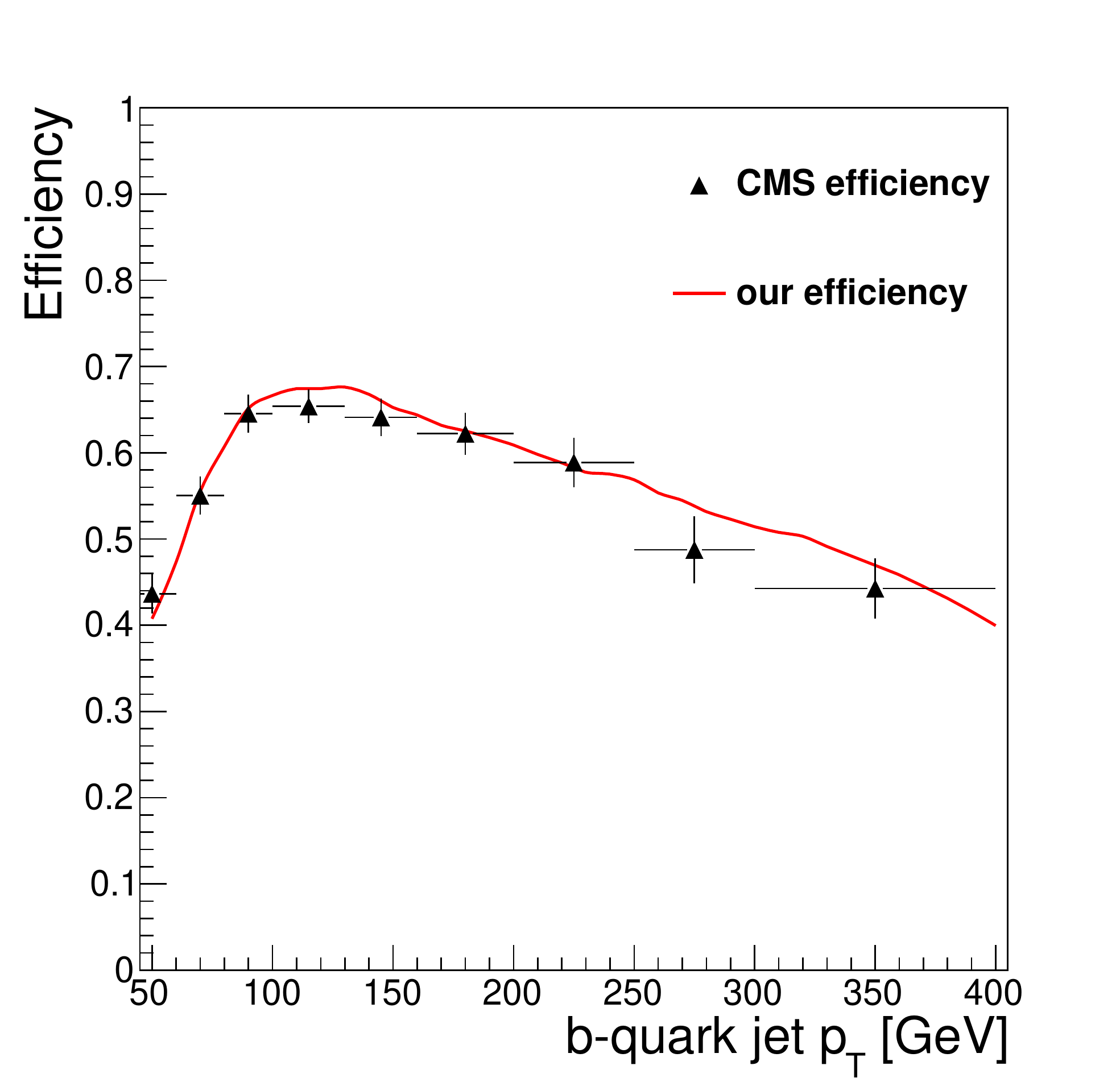}}
\caption{\it A comparison of the b-quark jet $p_{T}$ dependence of the b-tagging efficiency as reported by CMS (markers), with the corresponding distribution obtained within our framework using the DELPHES detector simulation (solid line).}\label{fig:btagvalidation}
\end{figure*}

Besides generally validating our framework for different production mechanisms and decay chains within the context of the CMSSM, it is also critical to perform a dedicated validation of the emulation of the b-tagging algorithm, which plays a crucial role in searches for third-generation squarks. For this, we revert to information provided in the right hand panel of Figure 2 of the same-sign dilepton search publication~\cite{Chatrchyan:2012sa}, which shows the jet transverse momentum dependence of the b-tagging efficiency as used by CMS. This is representative of all the b-tagging algorithms used in the searches we consider, as the efficiency and mis-tagging rates are very similar or identical to what is reported in this paper. We use this information to tune the corresponding b-tagging algorithm within the DELPHES detector simulation, and as Figure~\ref{fig:btagvalidation} shows, the jet transverse momentum dependent efficiency of the CMS b-tagging algorithms can therefore be well reproduced using our framework. A dedicated study of the lepton identification efficiency was also carried out and found to agree well with the information reported by CMS (from the left-hand panel of Figure 2 in~\cite{Chatrchyan:2012sa}), as confirmed by the results of Figure~\ref{fig:validation}.

Finally, in order to obtain a combined confidence level over all searches, we build a combined likelihood out of the individual likelihoods. By design of the individual topology searches, there is no event overlap, and hence we treat these searches as statistically independent from each other. Furthermore, since the information is not available, we apply no correlations for common systematic uncertainties. However, the typical size of these uncertainties when compared to those relevant for our analysis suggests that the final results of our studies are not impacted by neglecting these correlations. 

That being said, we would like to stress that the limits which will be quoted in this paper do not represent actual experimental limits. In order to achieve the most precise estimate of these limits, a careful consideration of all systematic uncertainties on the signal acceptance, and their corresponding correlations, need to be taken into account. This is a responsibility for the experimental collaborations and we are not trying to replace this important step. Nevertheless, we believe that the precision of the limits quoted in this paper is sufficient to establish meaningful mass limits relevant in the context of Natural-like SUSY spectra.

\section{Natural-like SUSY spectra}\label{sec:spectra}

The absence of any sign of SUSY at the LHC so far, when coupled with the discovery of a Higgs-like boson, has focussed the attention of both the experimental and theoretical communities on new benchmark scenarios defined in the context of Natural SUSY. However, in contrast to well defined models like the CMSSM, Natural SUSY models do not predict a unique sparticle spectrum. Instead, the term Natural SUSY seems to mainly refer to a class of SUSY models which are defined by a consideration of theoretical arguments relating to naturalness, as well as by current experimental constraints. On one hand, a Natural SUSY spectrum is motivated in order to avoid the current direct search constraints from the LHC, by decoupling the first and second generation squarks from the third-generation ones, while on the other hand, it is motivated from the theoretical perspective in the form of fine-tuning considerations arising from the desire to establish natural conditions for Electro Weak Symmetry Breaking in a supersymmetric theory i.e. without too much fine-tuning. Both of these aspects, however, are subjective and thus do not lead to a well-defined concept for establishing a unique Natural SUSY spectrum.

To overcome this ambiguity we establish a class of sparticle spectra, referred to as Natural-like spectra, which are inspired from the consideration of fine-tuning arguments, and comply with experimental search constraints. For the former, we are guided by the definition of \cite{Barbieri198863}, where a minimal natural SUSY model is typically characterised by at least three light third-generation squarks with masses well below 1 TeV, a gluino with a mass less than about 1.5 TeV, and light higgsinos of masses below 0.5 TeV. The stated mass ranges fulfill canonical fine-tuning criteria, but more importantly they also represent the current sensitivity range of direct SUSY searches at the LHC.    

\begin{figure*}[htb!]
\centering
\subfigure[]{\resizebox{6cm}{!}{\includegraphics{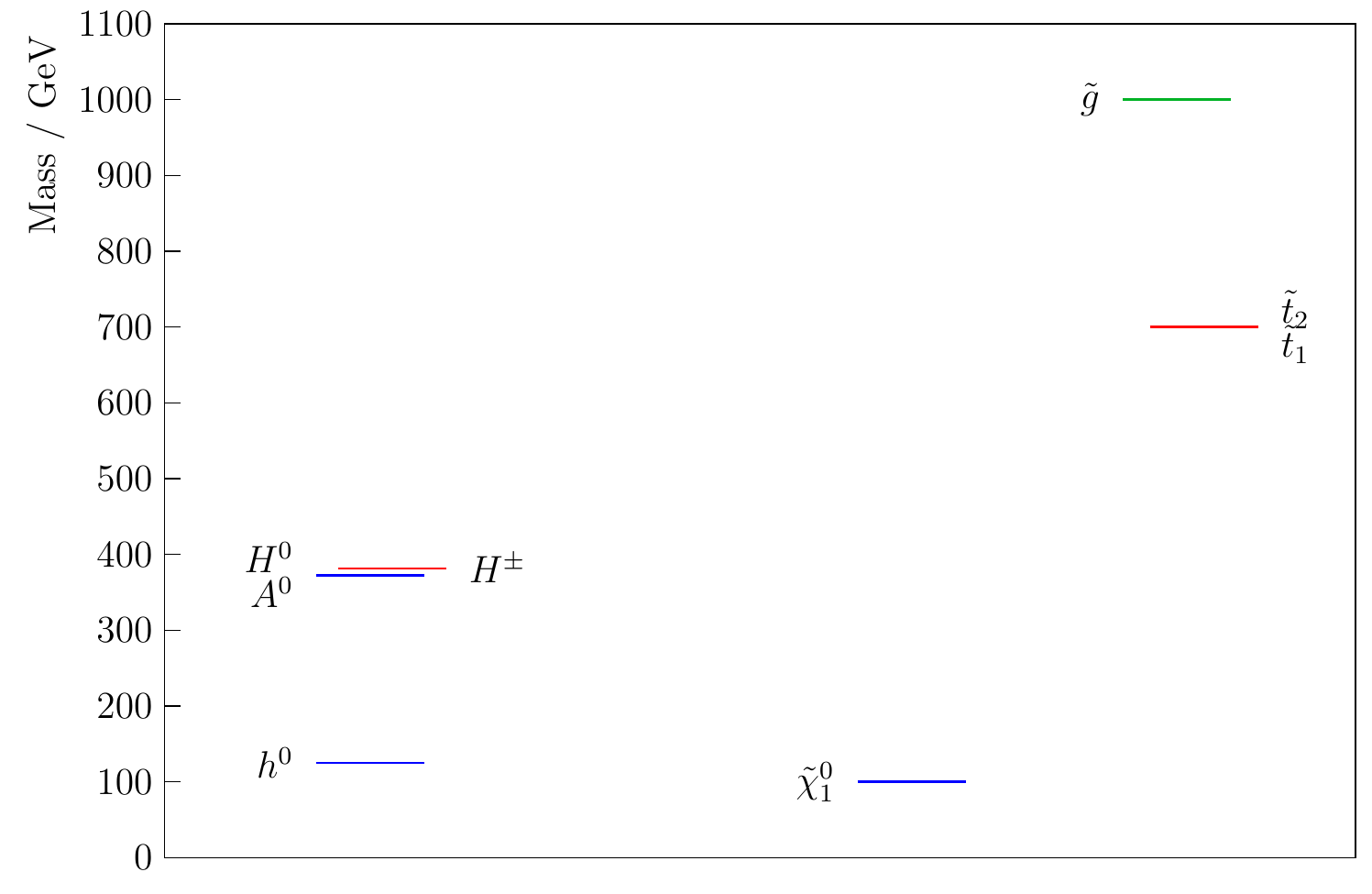}}\label{subfig:NS0}}
\subfigure[]{\resizebox{6cm}{!}{\includegraphics{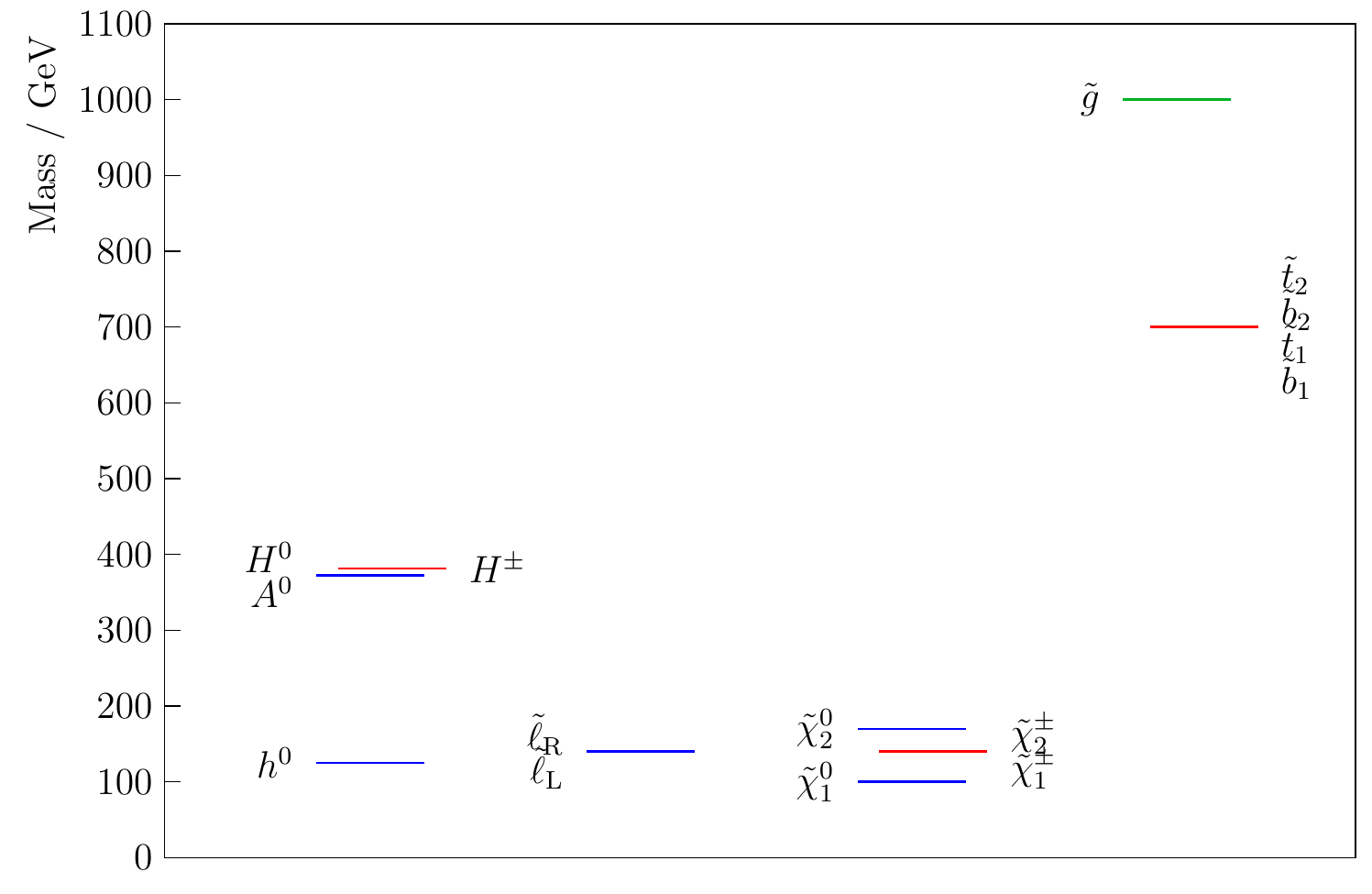}}\label{subfig:NS4}}
\caption{\it An illustration of the particle mass hierarchy for the Natural-like SUSY spectra NS0 (left) and NS4 (right).}\label{fig:NS04}
\end{figure*}

In the following, we define five Natural-like SUSY spectra (NS0-NS4) that are used to illustrate important properties of the inclusive SUSY topology searches, namely the number and type of leptons in the final state, and show how the dependence on this can be reduced when combining the most relevant inclusive searches. We also define $m_{\tilde{3G}}$ as the mass scale for all third-generation squarks in these spectra.
Table~\ref{tab:NSdef} defines the particle hierarchy of these benchmark points, where the gluino, third-generation squarks, and LSP are taken as free parameters in the studies that follow, and there is no requirement that $m_{\tilde{g}} > m_{\tilde{3G}}$, but simply that the gluino and third-generation squark masses are above the LSP. The mass differences, between the $\tilde{\chi}^{2}_{0}$ and $\tilde{\chi}^{1}_{0}$, and $\tilde{\chi}^{\pm}$/$\tilde{\ell}_{L,R}$ and $\tilde{\chi}^{1}_{0}$, are fixed for these illustrative spectra to $70$~GeV and $40$~GeV throughout respectively. The effect of breaking the mass degeneracies and varying the fixed mass splittings of the Natural-like spectra are discussed in the context of a systematic uncertainty in Section~\ref{sec:results}, where the impact on the definition of $m_{\tilde{3G}}$ is also discussed. The complexity in terms of sparticle content and relevant decay chains increases when going from NS0 to NS4. For illustration, the simplest (NS0) and most complicated (NS4) spectra are shown in Figure~\ref{fig:NS04} with some representative mass values chosen.

\begin{table*}[!tbh!]
\tbl{An overview of the sparticle content of the Natural-like SUSY spectra defined in this paper. The most important decay chains for each spectrum are also indicated.}
{
\begin{tabular}{| c | c | c | c | c | c |}
\hline
Spectra & NS0 & NS1 & NS2 & NS3 & NS4 \\
\hline
sparticle &
$\tilde{g}$ &
$\tilde{g}$ &
$\tilde{g}$ &
$\tilde{g}$ &
$\tilde{g}$ \\
content &
$\tilde{t_1},\tilde{t_2}$ &
$\tilde{t_1},\tilde{t_2},\tilde{b_1}$ &
$\tilde{t_1},\tilde{t_2},\tilde{b_1}$ &
$\tilde{t_1},\tilde{t_2},\tilde{b_1},\tilde{b_2}$ &
$\tilde{t_1},\tilde{t_2},\tilde{b_1},\tilde{b_2}$ \\
&
&
&
$\tilde{\chi}^{2}_{0}$ &
$\tilde{\chi}^{2}_{0}$ &
$\tilde{\chi}^{2}_{0}$ \\
&
&
&
$\tilde{\chi}^{\pm}$ &
$\tilde{\chi}^{\pm}$ &
$\tilde{\chi}^{\pm}, \tilde{\ell}_{L,R}$ \\
&
$\tilde{\chi}^{1}_{0}$ &
$\tilde{\chi}^{1}_{0}$ &
$\tilde{\chi}^{1}_{0}$ &
$\tilde{\chi}^{1}_{0}$ &
$\tilde{\chi}^{1}_{0}$ \\
\hline
main &
$\tilde{g} \to t\tilde{t}_{1,2}$ &
$\tilde{g} \to t\tilde{t}_{1,2}$, $b\tilde{b}_{1}$ &
$\tilde{g} \to t\tilde{t}_{1,2}$, $b\tilde{b}_{1}$ &
$\tilde{g} \to t\tilde{t}_{1,2}$, $b\tilde{b}_{1,2}$ &
$\tilde{g} \to t\tilde{t}_{1,2}$, $b\tilde{b}_{1,2}$ \\
decay &
$\tilde{t}_{1,2} \to t\tilde{\chi}^{1}_{0}$ &
$\tilde{t}_{1,2} \to t\tilde{\chi}^{1}_{0}$ &
$\tilde{t}_{1,2} \to t\tilde{\chi}^{1,2}_{0}$, $b\tilde{\chi}^{\pm}$ &
$\tilde{t}_{1,2} \to t\tilde{\chi}^{1,2}_{0}$, $b\tilde{\chi}^{\pm}$ &
$\tilde{t}_{1,2} \to t\tilde{\chi}^{1,2}_{0}$, $b\tilde{\chi}^{\pm}$ \\
chains &
&
$\tilde{b}_{1} \to b\tilde{\chi}^{1}_{0}$ &
$\tilde{b}_{1} \to b\tilde{\chi}^{2}_{0}$, $t\tilde{\chi}^{\pm}$ &
$\tilde{b}_{1,2} \to b\tilde{\chi}^{2}_{0}$, $t\tilde{\chi}^{\pm}$ &
$\tilde{b}_{1,2} \to b\tilde{\chi}^{2}_{0}$, $t\tilde{\chi}^{\pm}$ \\
&
&
&
$\tilde{\chi}^{\pm} \to W^{\pm} \tilde{\chi}^{1}_{0}$ &
$\tilde{\chi}^{\pm} \to W^{\pm} \tilde{\chi}^{1}_{0}$ &
$\tilde{\chi}^{\pm} \to W^{\pm} \tilde{\chi}^{1}_{0}$ \\
&
&
&
$\tilde{\chi}^{2}_{0} \to Z \tilde{\chi}^{1}_{0}$ &
$\tilde{\chi}^{2}_{0} \to Z \tilde{\chi}^{1}_{0}$ &
$\tilde{\chi}^{2}_{0} \to Z \tilde{\chi}^{1}_{0}$, $\tilde{\ell}\ell$ \\
&
&
&
&
&
$\tilde{\ell} \to \ell \tilde{\chi}^{1}_{0}$ \\
\hline
\end{tabular}
\label{tab:NSdef}}
\end{table*}

To minimise the impact of statistical uncertainties in our work, for each of the results reported in the following, we generate signal events corresponding to at least $200\textrm{fb}^{-1}$ of data, and normalise the signal expectations to those as reported in the publications. 

\begin{figure*}[htb!]
\centering
\resizebox{12cm}{!}{\includegraphics{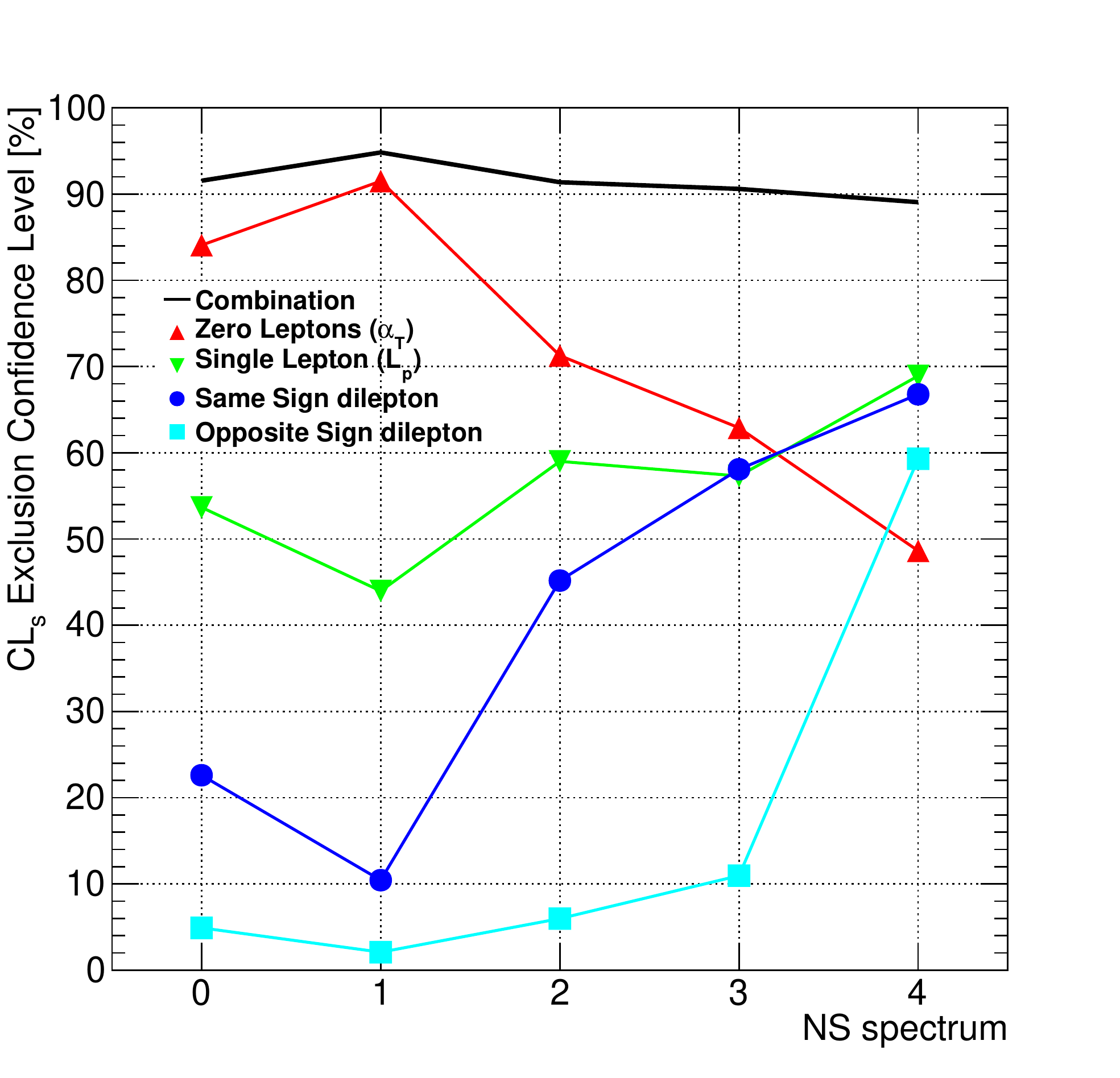}}
\caption{\it Determination of the $CL_{s}$ exclusion confidence level for the set of NS spectra defined in this paper, in which the gluino, third-generation squarks and LSP masses are fixed to the representative values of $1000$~GeV, $700$~GeV and $100$~GeV, respectively. The $CL_{s}$ value is plotted as a function of increasing underlying complexity (NS0 to NS4), shown for the individual searches, and in combination.}\label{fig:cls_vs_complexity}
\end{figure*}

In order to determine the importance of combining relevant topology searches, we first perform a calculation of the $CL_{s}$ exclusion value for a set of NS spectra in which the gluino, third-generation squarks and LSP masses are fixed to the representative values shown in Figure~\ref{fig:NS04}, of $1000$~GeV, $700$~GeV and $100$~GeV, respectively. The results of this study are shown in Figure~\ref{fig:cls_vs_complexity}. The calculated $CL_{s}$ value for individual searches varies strongly as the level of complexity increases from NS0 to NS4 (i.e. going from left to right in the plot). While the zero-lepton $\alpha_{T}$ search dominates the combined exclusion confidence for the simple spectra NS0 and NS1, decay chains producing leptonic final states become more important as the complexity of the underlying spectrum increases (see Table~\ref{tab:NSdef}). Only when combining all individual searches does the $CL_{s}$ value remain stable as a function of the underlying spectrum complexity.

\begin{figure*}[htb!]
\centering
\subfigure[]{\resizebox{6cm}{!}{\includegraphics{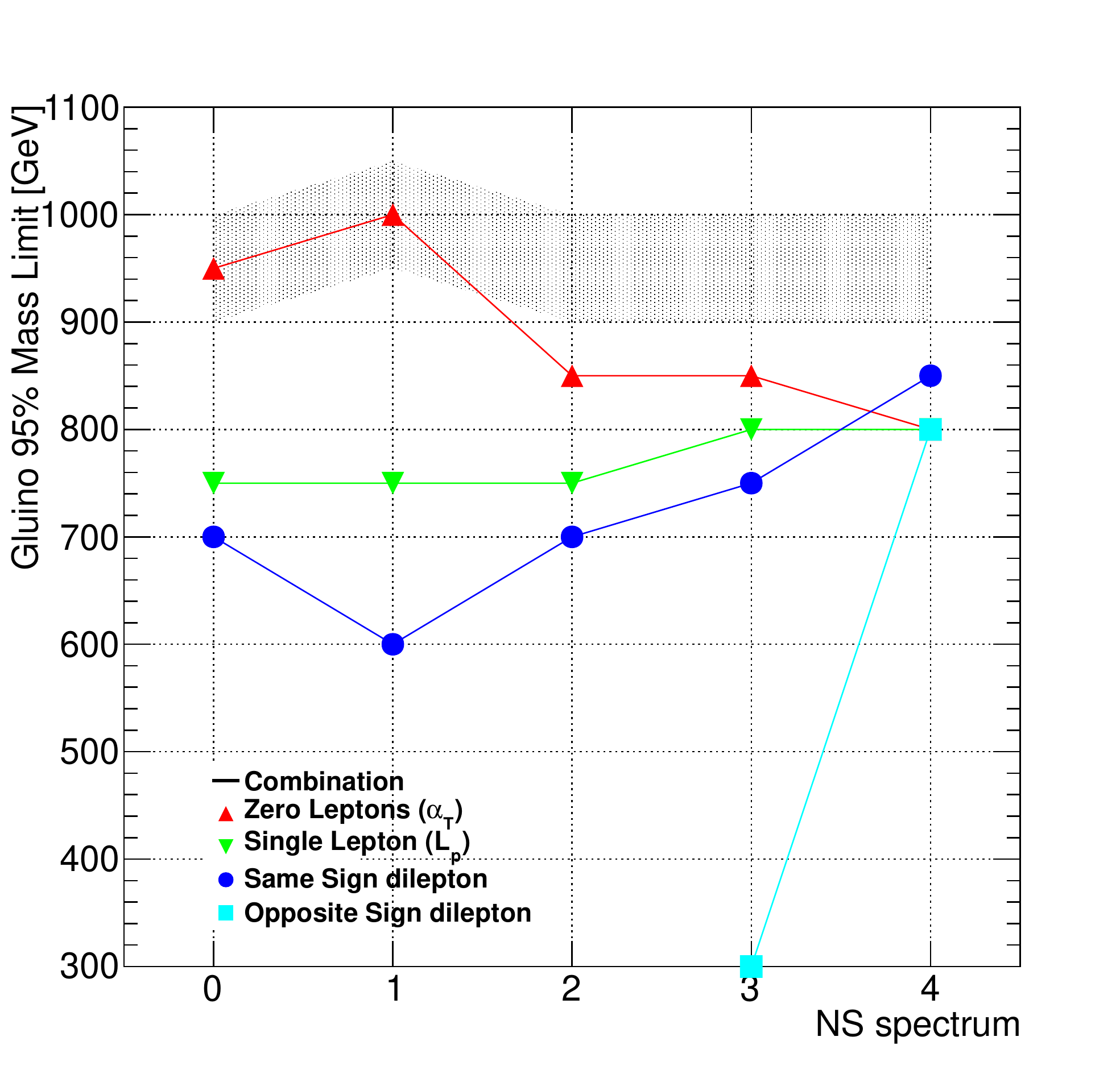}}\label{subfig:mgluino_vs_complexity}}
\subfigure[]{\resizebox{6cm}{!}{\includegraphics{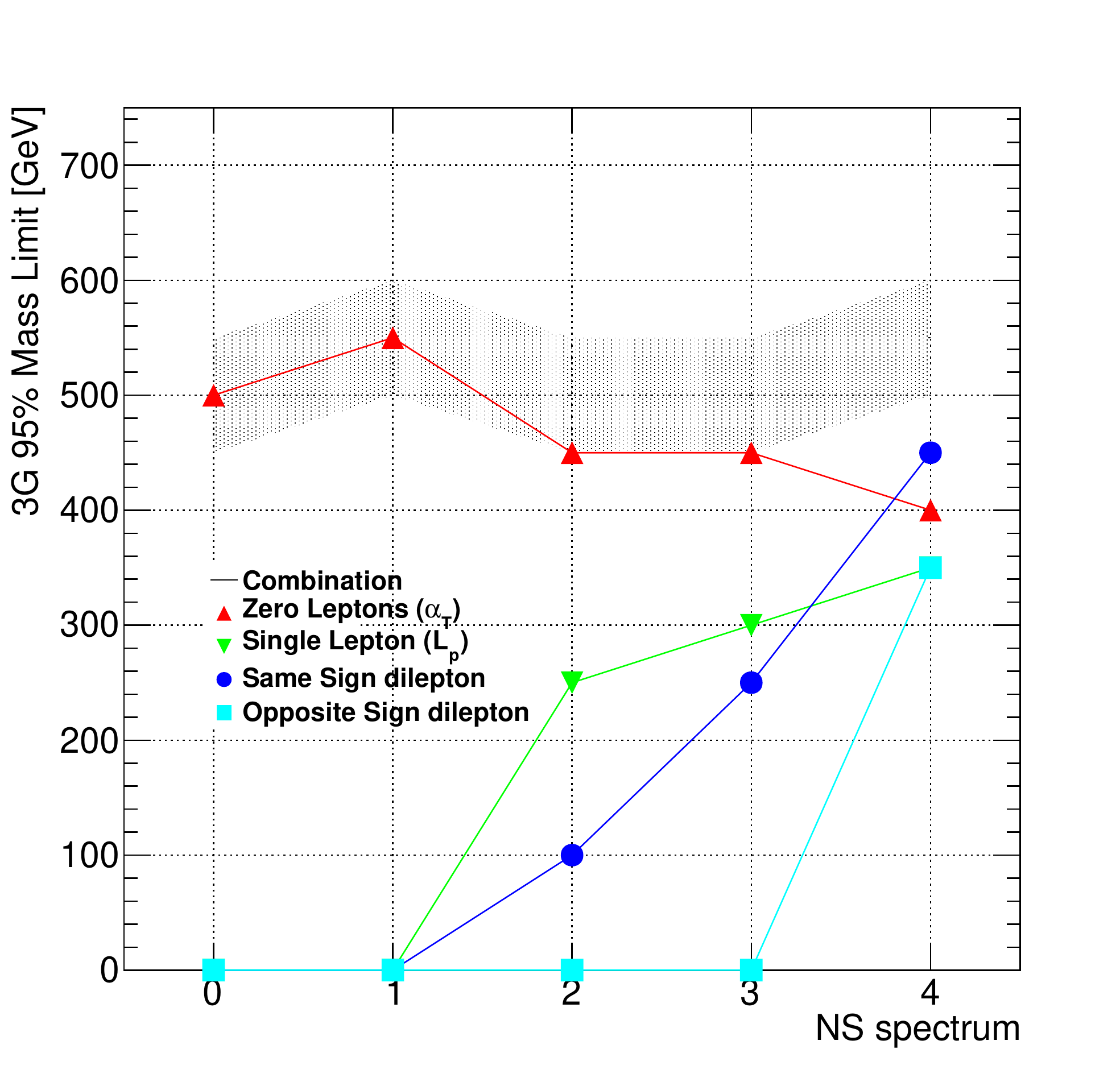}}
\label{subfig:mthreeG_vs_complexity}}
\caption{\it The gluino (left) and third-generation squark (right) mass limits for the set of NS spectra defined in this paper, shown for both the individual searches, and in combination. The grey band around the combined limits represents the uncertainty to which the combination was determined.}\label{fig:mass_vs_complexity}
\end{figure*}

As a next step, we perform a scan in the gluino and third-generation squark mass plane, for a fixed LSP mass of $100$~GeV, for each of the Natural SUSY spectra NS0-NS4. Based on this scan, we determine the 95\% $CL_{s}$ exclusion mass limits for the gluino $m_{\tilde{g}}$, or third-generation squarks $m_{\tilde{3G}}$. The results of this scan are summarised in Figure~\ref{fig:mass_vs_complexity}. The mass limits shown represent the cases where the gluino, Figure~\ref{subfig:mgluino_vs_complexity}, or third-generation squark mass, Figure~\ref{subfig:mthreeG_vs_complexity}, is ruled out, irrespective of other masses in the spectra. The size of the shaded band on the combination represents the $50$~GeV granularity of our scan, which dominates in this particular case over all other systematic uncertainties. As already demonstrated in Figure~\ref{fig:cls_vs_complexity}, the relative impact of individual searches depends strongly on the underlying sparticle content considered, whereas the combination of searches is relatively stable. While the individual mass limits, both for $m_{\tilde{g}}$ as well as $m_{\tilde{3G}}$, can vary by several hundred GeV between NS0 and NS4, the combined mass limit is stable, within the uncertainty, at around $950$~GeV and $500$~GeV for $m_{\tilde{g}}$ and $m_{\tilde{3G}}$, respectively. Therefore, for the class of spectra considered here, these mass limits represent universal limits on these quantities in the absence of systematic uncertainties, which are discussed in Section~\ref{sec:results}.

This reinforces the importance of combining relevant topology searches in order to make the most model independent interpretations, and is one of the important conclusions of this paper.

\section{Limits on gluino and third-generation squark masses in the context of Natural-like SUSY spectra}\label{sec:results}
As discussed in Section~\ref{sec:spectra}, the combination of relevant topology searches is crucial when the most model independent interpretation of results is required. This section extends on this result by addressing the question on what current experimental search results say on the masses of the gluino and third-generation squarks in the context of Natural-like SUSY models. 
For this purpose, we now only consider searches in combination, and take the most complicated spectrum, NS4, as our benchmark scenario. We also extend our analysis with a more detailed study of systematic uncertainties related to breaking the mass degeneracies and to changes of the underlying spectrum complexity. Furthermore, a scan over the LSP mass is undertaken to determine the dependence of the gluino and third-generation squark mass limits on this important quantity.

\subsection{Results based on 2011 searches}\label{sec:results2011}  

\begin{figure*}[htb!]
\centering
\subfigure[]{\resizebox{6cm}{!}{\includegraphics{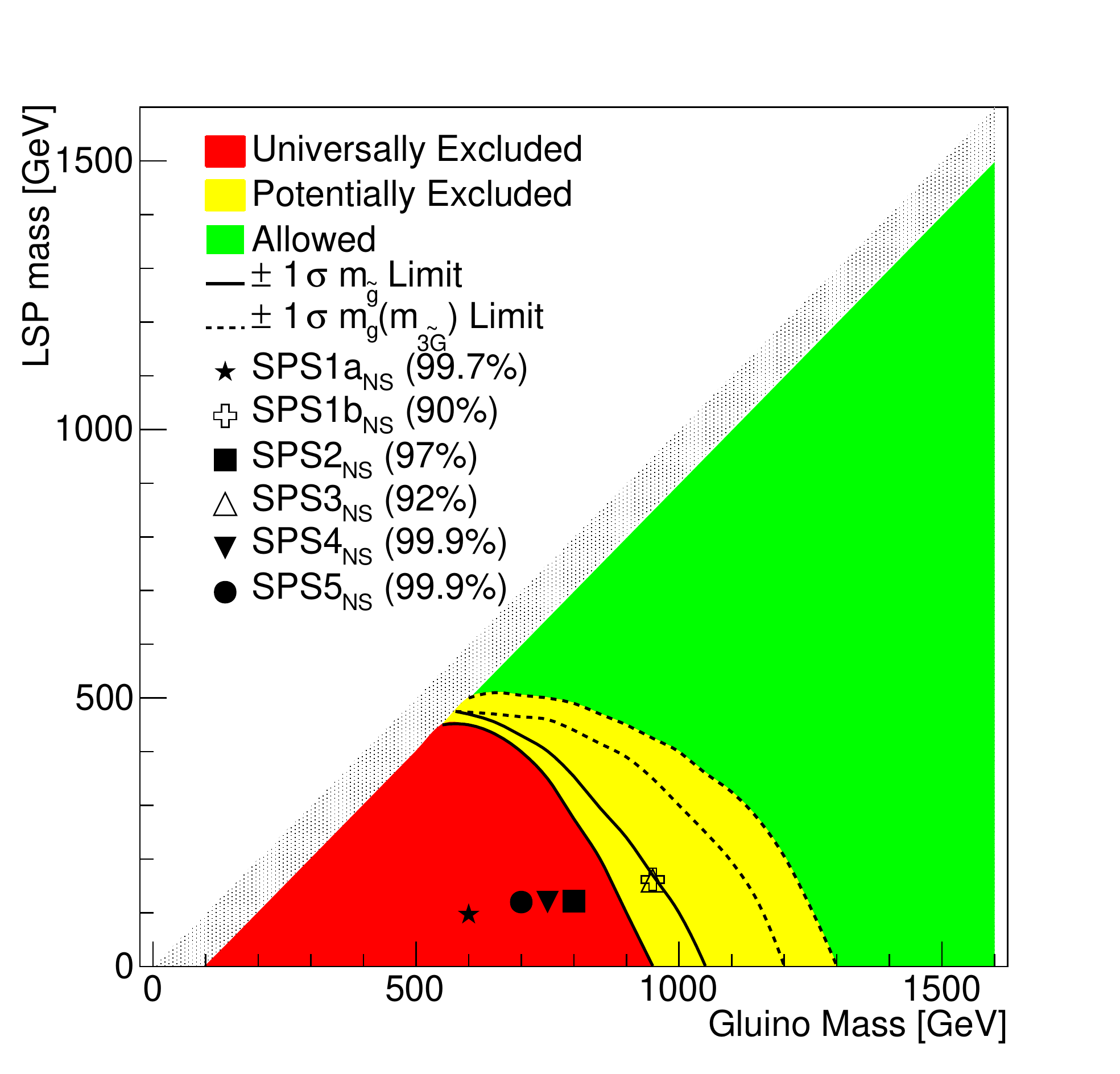}}
\label{subfig:noscanspsgluino}
}
\subfigure[]{\resizebox{6cm}{!}{\includegraphics{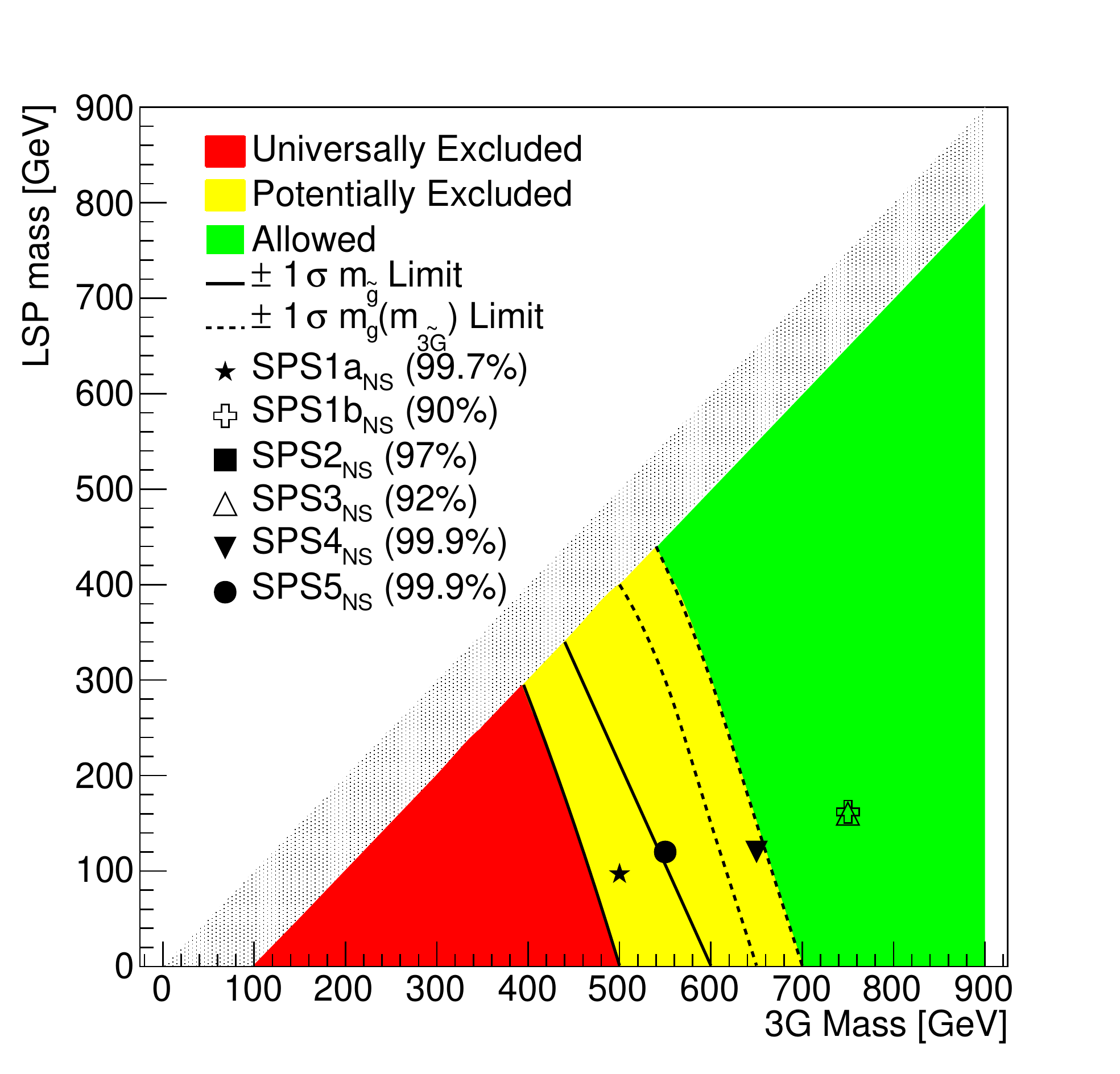}}
\label{subfig:noscansps3g}
}
\subfigure[]{\resizebox{6cm}{!}{\includegraphics{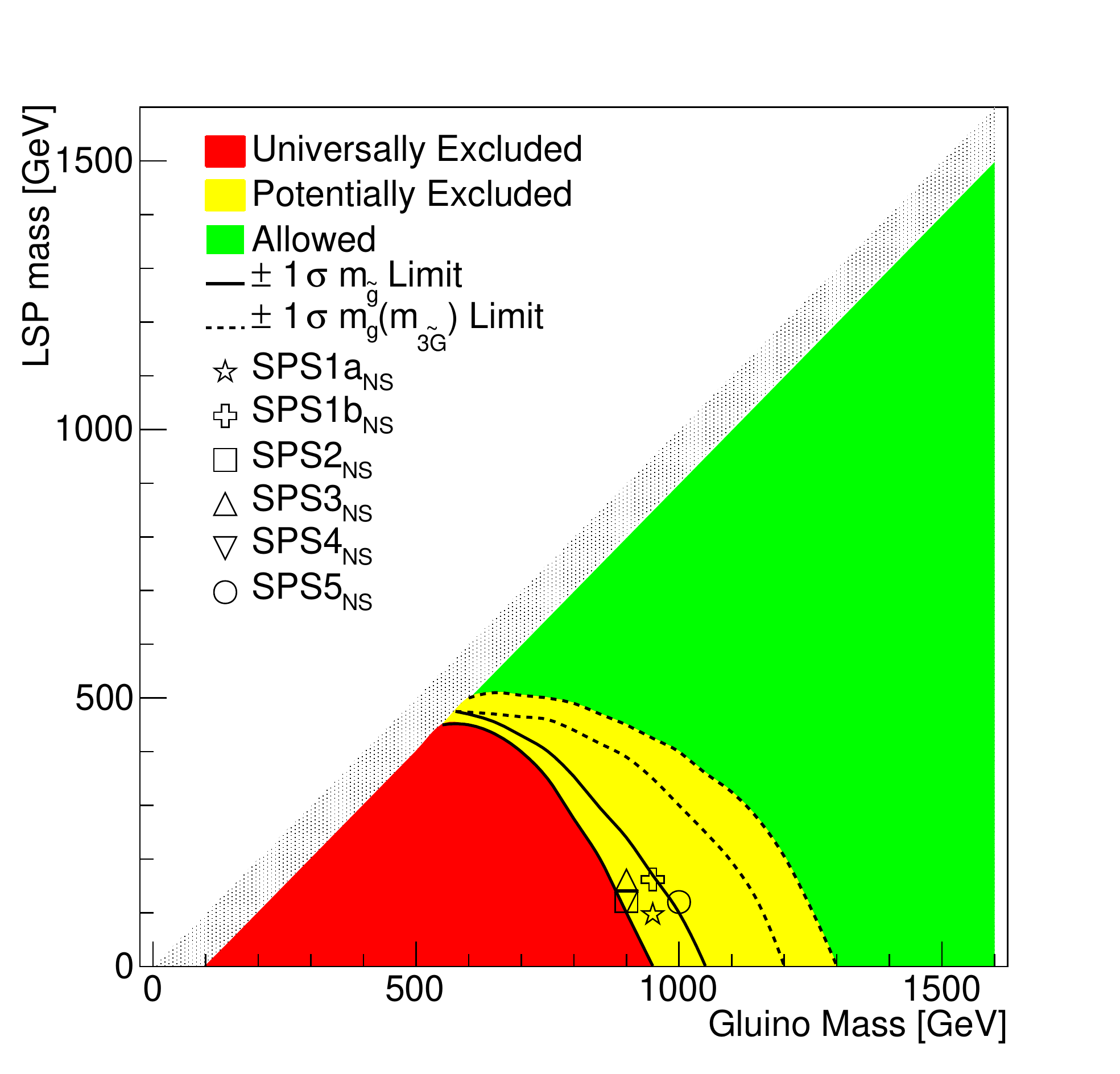}}
\label{subfig:scanspsgluino}
}
\subfigure[]{\resizebox{6cm}{!}{\includegraphics{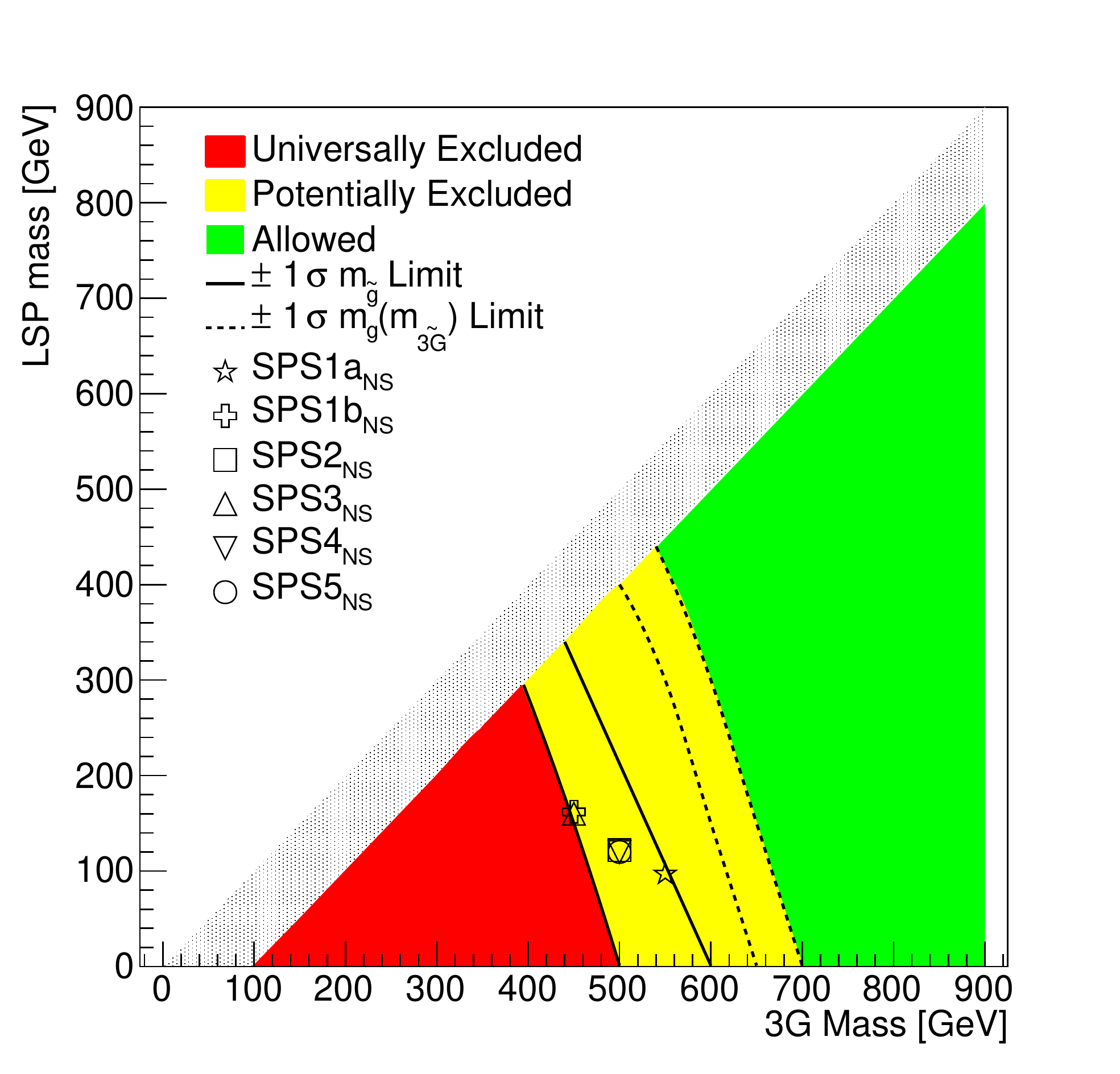}}
\label{subfig:scansps3g}
}
\caption{\it Gluino and third-generation limit plots for the 2011 searches. For further details on the colour scheme used, see the text in Section~\ref{sec:results2011}. Top: The distributions of the SPS$_{NS}$ benchmark scenarios in the $(m_{\tilde{g}}, m_{LSP})$ plane (left) and $(m_{\tilde{3G}}, m_{LSP})$ plane (right). Bottom: The 95\% mass limits obtained by scanning the gluino (left) and third-generation squark (right) mass planes for each of the SPS$_{NS}$ benchmark scenarios.}\label{fig:spsvalidation}
\end{figure*}

As in Section~\ref{sec:spectra}, we compute universal mass limits on the gluino and third-generation squark masses irrespective of other particles in the spectrum, as a function of the LSP mass. We present our results in the mass plane of the LSP versus the gluino and third-generation squarks separately, keeping the format the same as those presented in the SMS scenarios by the CMS and ATLAS experiments, to allow for easier comparison of limits. We do not provide an interpretation for the region where the mass splitting between the gluino or third-generation squark mass and the LSP is less than $100$~GeV. This region is subject to significant experimental uncertainties relating to signal acceptance, as well as theoretical uncertainties arising from the modelling of initial state radiation, which plays an important role for very compressed spectra. The results of this scan for the combined 2011 $\sqrt{s}=7$~TeV CMS searches corresponding to $5\textrm{fb}^{-1}$ are shown in Figure~\ref{fig:spsvalidation} for the gluino (left hand panels) and third-generation squark masses (right hand panels).

The shaded area represents the region of compressed spectra for which we do not provide an interpretation. The solid black lines define the $\pm 1 \sigma$ region around the 95\% $CL_{s}$ exclusion limit on the gluino or third-generation squark masses, which are independent of the contribution from other sparticle production mechanisms. Therefore, these limits do not depend on the details of the underlying sparticle content of a given Natural-like SUSY spectrum and thus in the following are referred to as universal limits on the gluino and third-generation squark masses. The $\pm 1 \sigma$ region is defined by systematic uncertainties arising from (a) the uncertainty on the definition of the $CL_{s}$ statistic of 95\%$^{+3\%}_{-7\%}$, (b) the granularity of our scan in the $(m_{\tilde{g}}, m_{LSP})$ and $(m_{3G}, m_{LSP})$ planes, and (c) the effect of varying the underlying spectrum assumptions of NS4. For (a) and (b), this typically translates into an uncertainty of around $\pm 25$~GeV each on the mass limits. For (c), we vary systematically the mass splittings between the $\tilde{t}_{1,2}$ and $\tilde{b}_{1,2}$, the $\tilde{\chi}^{\pm}$ and $\tilde{\ell}_{L,R}$, as well as between the $\tilde{\chi}^{2}_{0}$, $\tilde{\chi}^{\pm}$ and $\tilde{\chi}^{1}_{0}$. This comprehensive variation of the underlying spectrum assumptions allows different decay chains to contribute and therefore changes the signal acceptance of the individual topology searches. This enables us to obtain an estimate of how well NS4 represents the class of Natural-like SUSY spectra. Depending on the chosen configuration of the underlying spectrum, as well as on the mass of the LSP, this uncertainty, which gauges the model-independence of our limits, yields values in the range of $\pm 25$~GeV to $\pm 50$~GeV on the mass limits. All these systematic uncertainties are combined in quadrature and are taken into account in the final results. As can be seen in Figure~\ref{fig:spsvalidation}, the consideration of these systematic uncertainties results in $\pm 1 \sigma$ regions around the final universal limit of up to $100$~GeV on the gluino and third-generation squark masses, dependent on the mass of the LSP. 

Analogously, the black dotted lines define the $\pm 1 \sigma$ region in which the gluino (third-generation squark) mass might be excluded, when considering third-generation squark (gluino) masses above the threshold imposed by their universal exclusion limit (i.e. the $\pm 1 \sigma$ region defined by the solid black lines). In other words, in contrast to the universal limit defined above, this limit allows for additional production mechanisms to contribute to the final limit.  Although we allow for contributions from $\tilde{g}-\tilde{3G}$ production as well as direct higgsino and slepton production in our analysis, these processes have no significant impact on the final results. In the case of $\tilde{g}-\tilde{3G}$ production, the overall cross section is much smaller than those for gluino and third-generation squark pair production, while for direct higgsino and slepton production, the signal acceptance of the inclusive searches for the direct production of these sparticles is negligible. Therefore, in practice, the only relevant additional contributions to the gluino mass limit stems from direct production of third-generation squarks and vice versa. For example, for a given LSP mass, the maximal exclusion of the gluino mass, represented by the  $\pm 1 \sigma$ region defined by the black dotted lines, is accomplished by moving $m_{\tilde{3G}}$ just above its universal limit, which is represented by the $\pm 1 \sigma$ region defined by the solid black lines. In this configuration, direct third-generation squark pair production can contribute maximally, yielding the largest overall signal yield for the topology searches. Therefore, this provides the most stringent, but spectrum dependent, limit on the gluino mass. Conversely, the best limit on $m_{\tilde{3G}}$ is obtained when the maximally allowed production cross section for gluino pair production is reached.
Care must be taken when defining $m_{\tilde{3G}}$ in the case of large mass splittings between the third generation squarks, specifically in the scenario where not all the third generation squarks are above or below their respective $-1\sigma$ conditional limit. This is because squark masses above this limit have no measurable contribution to the searches, due to their small cross-section, and hence must not be allowed to affect the value of $m_{\tilde{3G}}$ disproportionately. In this scenario, any third-generation squark masses above their $-1\sigma$ conditional limit are capped at this value, and the numerical average calculated accordingly. For example, if two of the four third-generation squark masses are above their conditional limit, the numerical average is calculated with these two masses capped at the conditional limit. Otherwise, if all third-generation squarks are either above or below this limit, the numerical average is calculated as normal. For a future publication, we are considering alternative definitions of $m_{\tilde{3G}}$, which directly take into account the cross-section contribution of each individual squark. These refined definitions are expected to further reduce the spectrum dependent systematic uncertainties currently arising from the straightforward and simple definition of $m_{\tilde{3G}}$, especially for spectra with large mass splittings, thus making the universal limits on the third generation squark masses even more precise.

Based on this categorisation, we define three regions in our final limit plots, which we correspondingly refer to as ``traffic light'' plots. These plots facilitate a very simple interpretation of our results and allow for a straightforward interpretation of any Natural-like SUSY spectrum with our limits. As can be seen in Figure~\ref{fig:spsvalidation}, the first region, referred to as universally excluded (red area), reaches up to the $-1 \sigma$ line of the universal limit. We have chosen the $-1 \sigma$ exclusion contour in order to establish a conservative definition of this exclusion region. Therefore, a given Natural-like SUSY spectrum is universally excluded at the 95\% confidence level if either its gluino mass {\it or} average third-generation squark mass is located in the corresponding red area of our ``traffic light'' plots.

The second region (yellow area) extends from the $-1 \sigma$ line of the universal limit to the $+1 \sigma$ line of the spectrum dependent limit. A given Natural-like SUSY spectrum may or may not be excluded at the 95\% confidence level if both its gluino {\it and} average third-generation squark masses are located in the corresponding yellow area of our ``traffic light'' plots. In this case, the exclusion confidence from the combination of topology searches will depend on the details of the spectrum, and can only be determined with a dedicated calculation.

The third region (green area) is bounded from below by the $+1 \sigma$ line of the spectrum dependent limit. A given Natural-like SUSY spectrum is not excluded at the 95\% confidence level if either its gluino mass or average third-generation squark mass is located in the corresponding green area of our ``traffic light'' plots, and neither are located in the universally excluded region (red area). 

With this three-area approach matching the colour code of a traffic light, we have established a very convenient and simple way to confront a given Natural-like SUSY spectrum with our limits. In the case where both the gluino and third-generation squark masses are located in the yellow region of our limit plots, a spectrum dependent calculation of the $CL_{s}$ value is required. Otherwise, the appropriate conclusions may be drawn.

Based on the analysis of the 2011 searches, we find that for low LSP masses, gluino masses up to $900$~GeV are excluded, reducing to about $550$~GeV by an LSP mass of $450$~GeV. When considering potential contributions from third-generation squark pair production, these limits increase to $1250$~GeV at low LSP masses, and to about $600$~GeV by an LSP mass of $500$~GeV. For the third-generation squarks, masses of up to $475$~GeV are excluded at low LSP masses, falling to about $400$~GeV by an LSP mass of $300$~GeV. Again, when considering potential contributions from gluino pair production, this limit increases to about $675$~GeV for low LSP masses in the spectrum dependent case, and falls to $550$~GeV by an LSP mass of $450$~GeV. A summary of these exclusion limits are provided in the upper third of Table~\ref{tab:limits}.

\subsection{Verification of limit universality using historical SPS SUSY benchmark points}
\begin{figure*}[htb!]
\centering
\subfigure[]{\resizebox{6cm}{!}{\includegraphics{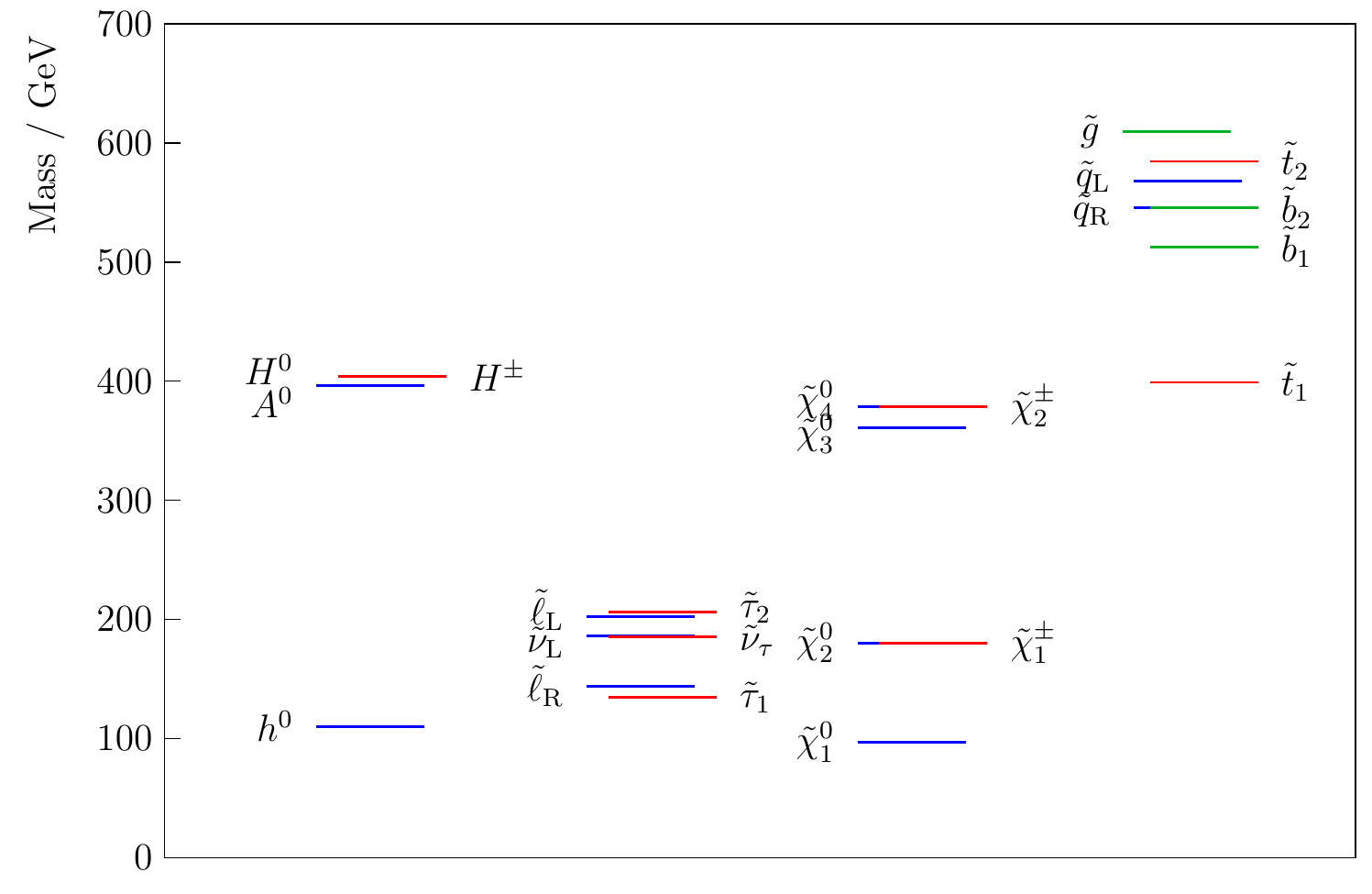}}\label{subfig:sps1a}}
\subfigure[]{\resizebox{6cm}{!}{\includegraphics{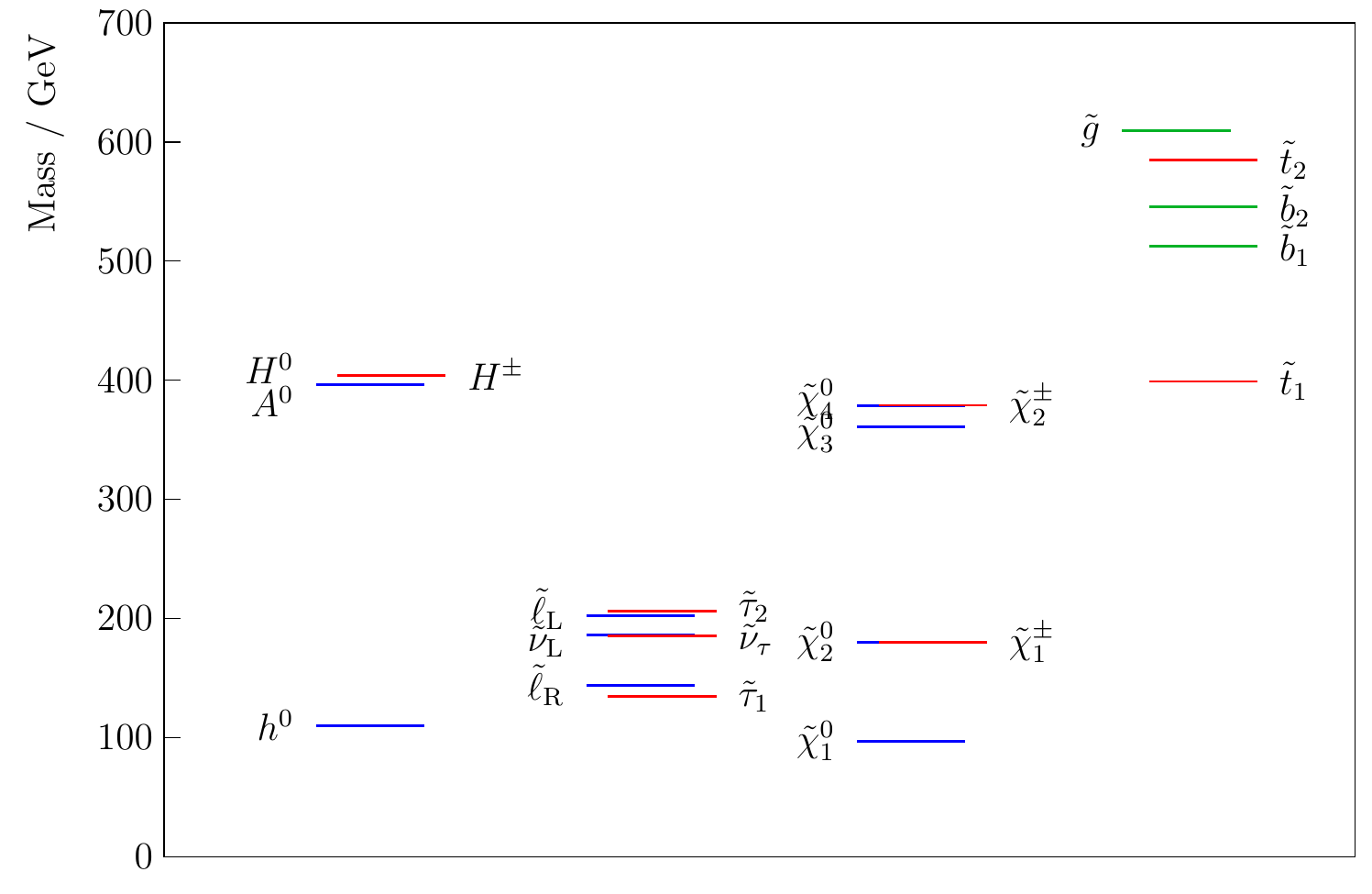}}\label{subfig:spsns1a}}
\caption{\it Comparison of the SPS1a spectrum before (left) and after (right) the first and second generation squarks are moved to the multi-TeV scale. The spectrum shown in the right-hand panel defines SPS1a$_{NS}$ in the context of Natural-like SUSY spectra.}\label{fig:sps1a3G}
\end{figure*}

\begin{figure*}[htb!]
\centering
\subfigure[]{\resizebox{6cm}{!}{\includegraphics{spsns1a}}
\label{subfig:sps1a2}
}
\subfigure[]{\resizebox{6cm}{!}{\includegraphics{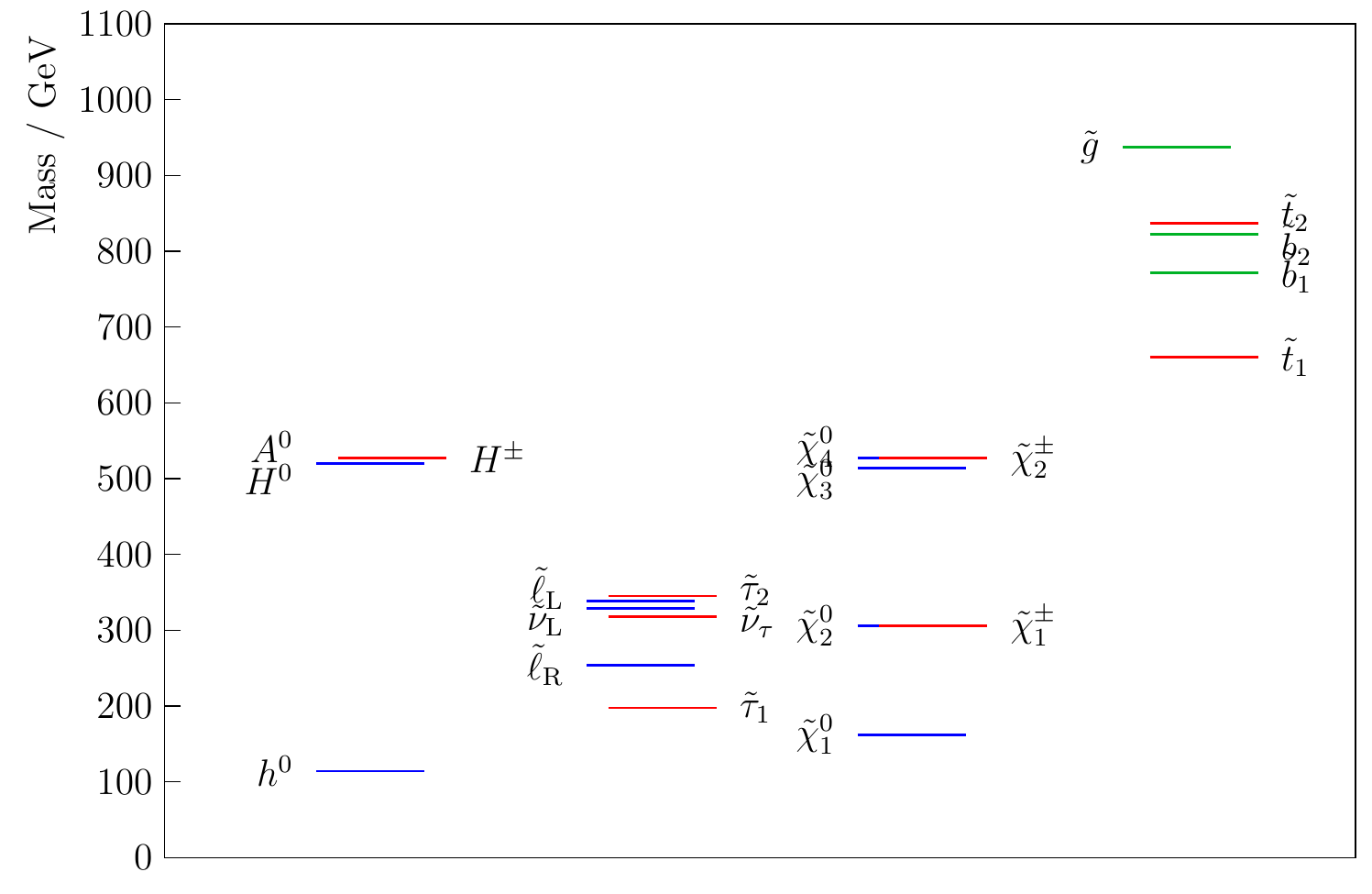}}
\label{subfig:sps1b}
}
\subfigure[]{\resizebox{6cm}{!}{\includegraphics{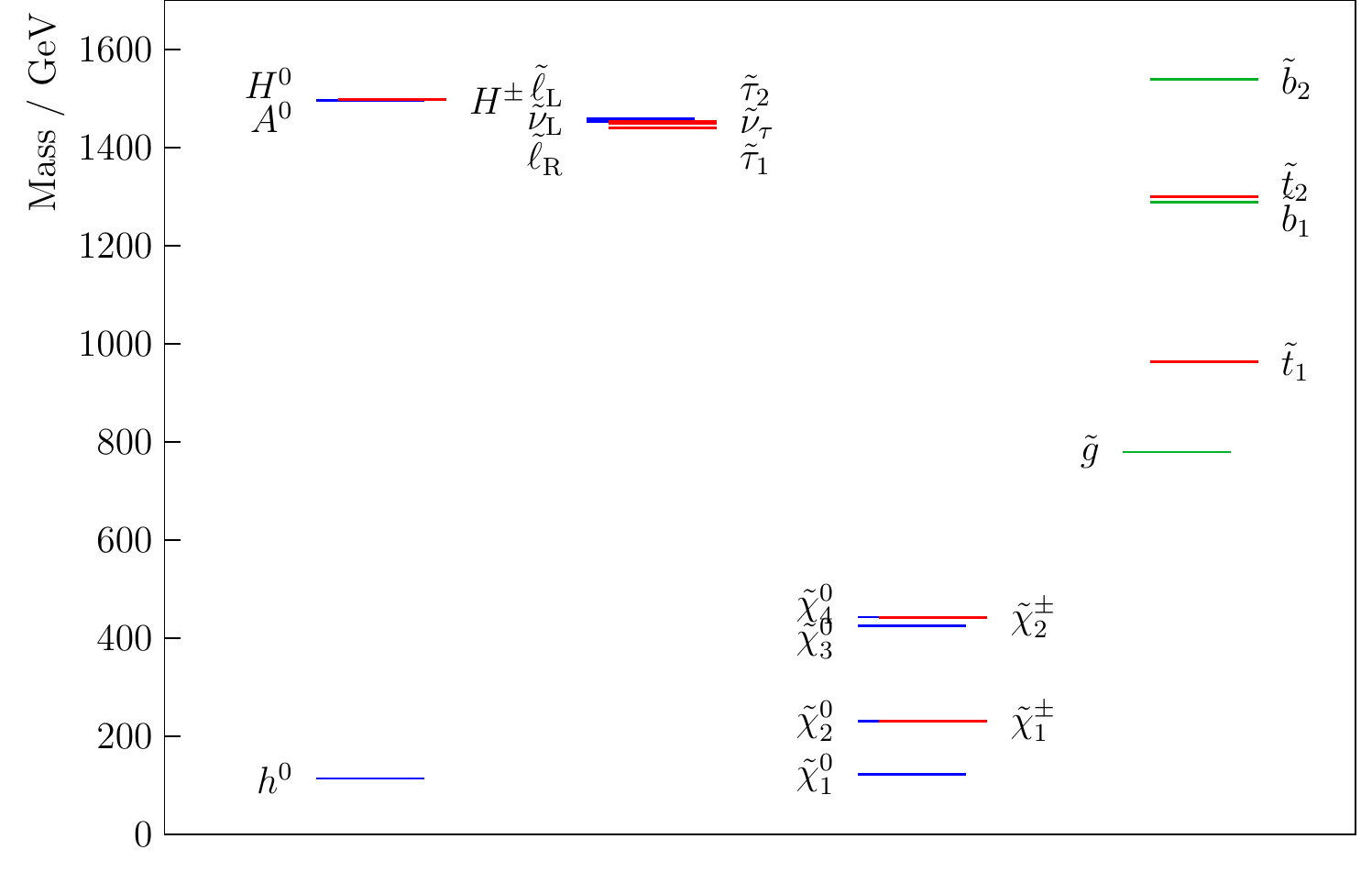}}
\label{subfig:sps2}
}
\subfigure[]{\resizebox{6cm}{!}{\includegraphics{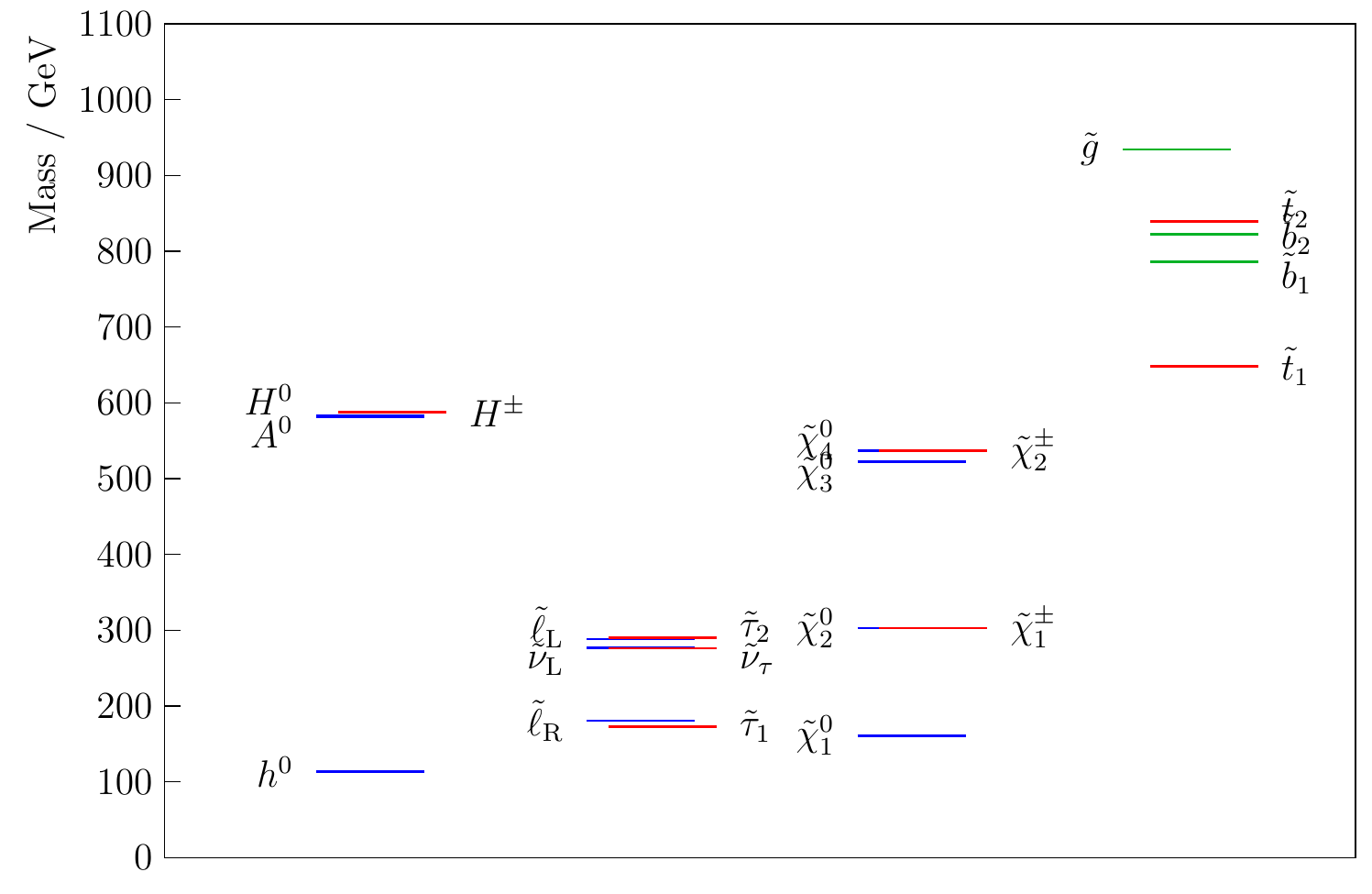}}
\label{subfig:sps3}
}
\subfigure[]{\resizebox{6cm}{!}{\includegraphics{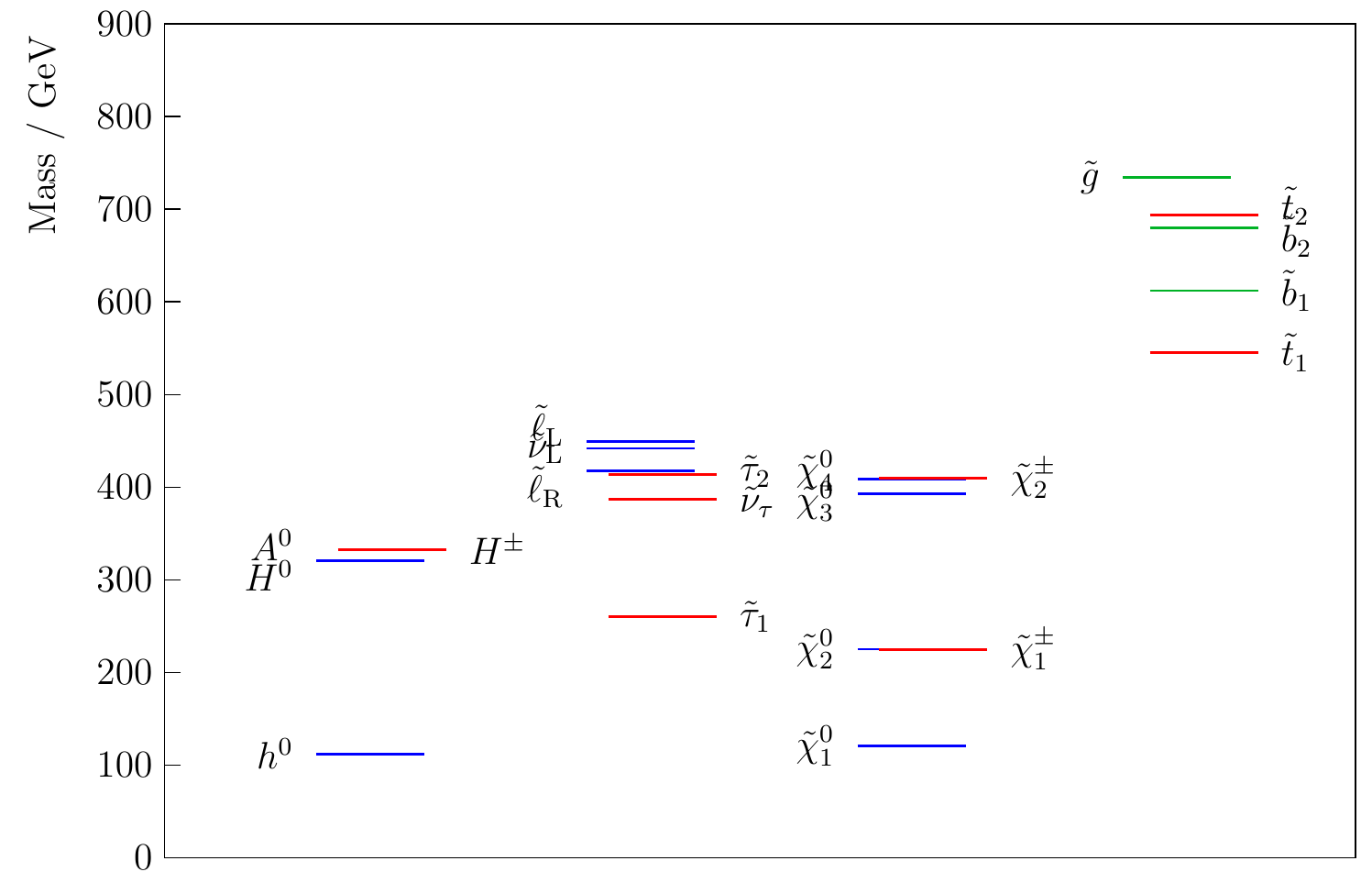}}
\label{subfig:sps4}
}
\subfigure[]{\resizebox{6cm}{!}{\includegraphics{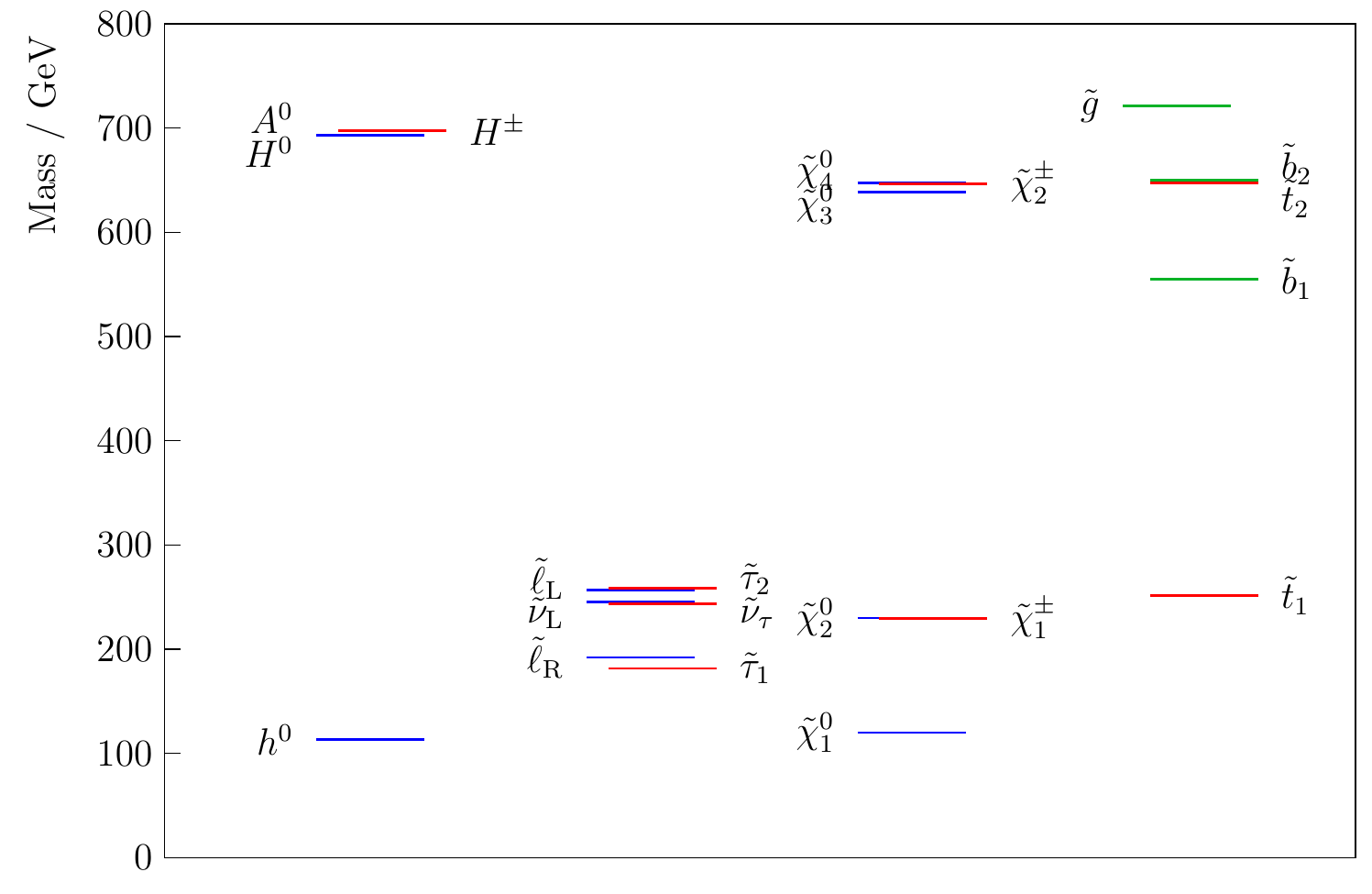}}
\label{subfig:sps5}
}

\caption{\it Illustrations of the SPS$_{NS}$ Natural-like SUSY spectra used in our analysis. These spectra are obtained by transforming the original SPS benchmark points, by setting the first and second generation squark masses at the multi-TeV scale. Figures~\ref{subfig:sps1a2} to~\ref{subfig:sps5} show SPS1a$_{NS}$ followed by SPS1b$_{NS}$ through to SPS5$_{NS}$.}\label{fig:spsNS}
\end{figure*}

To demonstrate the concept of the model-independence of our mass limits, we refer to the historical SPS~\cite{springerlink:10.1007/s10052-002-0949-3} SUSY benchmark points, as originally motivated in the context of the CMSSM. Specifically, we consider the SPS1a to SPS5 benchmark points as defined in~\cite{springerlink:10.1007/s10052-002-0949-3}. By setting the first and second generation squark masses at the multi-TeV scale, we are able to interpret these complete and complex spectra in the context of Natural-like scenarios. We refer to these modified benchmark points as $\textrm{SPS1a}_{NS}$ to $\textrm{SPS5}_{NS}$. Figure~\ref{fig:sps1a3G} shows the spectrum of the famous SPS1a benchmark point before and after the transformation to $\textrm{SPS1a}_{NS}$. All SPS$_{NS}$ spectra considered in this analysis are displayed in Figure~\ref{fig:spsNS}. These spectra are based on a complete SUSY model and thus exhibit a large variation in the underlying complexity.

First, we calculate for each of these redefined SPS points, the $CL_{s}$ exclusion confidence level for the combination of all searches considered in our analysis. The top panels of Figure~\ref{fig:spsvalidation} shows the distribution of these $SPS_{NS}$ spectra in the $(m_{\tilde{g}}, m_{LSP})$ and $(m_{\tilde{3G}}, m_{LSP})$ planes. The solid markers represent exclusion confidence values above 95\%, while the hollow markers represent values below 95\%. For SPS1a$_{NS}$, we calculate an average third-generation squark mass of $m_{\tilde{3G}}=$ 515 GeV. As can be seen from Figure~\ref{subfig:noscansps3g}, this value resides in the model-dependent exclusion region (yellow area) of our 2011 ``traffic light'' plot, indicating that this configuration is not ruled out {\it per se}, but could be ruled out, depending on contributions from gluino-gluino production. Since the gluino mass for the SPS1a$_{NS}$ point is $m_{\tilde{g}}=$ 610 GeV, and is located in the universal exclusion region (red area) of our 2011 limits (see Figure~\ref{subfig:noscanspsgluino}), our 2011 ``traffic light'' plot predicts that SPS1a$_{NS}$ is well excluded by the combination of the 2011 searches. Calculating the $CL_{s}$ exclusion confidence level for this spectrum using our framework yields a value of 99.7\%, confirming that this spectrum is indeed excluded. However, with an average third-generation mass of $m_{\tilde{3G}}=$ 770 GeV and a gluino mass of $m_{\tilde{g}}=$ 940 GeV, the SPS1b$_{NS}$ spectrum resides in the non-excluded region (green area) and model-dependent exclusion region (yellow area) for third-generation squark and gluino masses, respectively. Therefore, based on the interpretation of the ``traffic light'' plot, we would expect that this spectrum is not ruled out by the 2011 combination. This is confirmed by a dedicated full calculation of the corresponding $CL_{s}$ exclusion confidence level, which is determined to be 90\% and hence not excluded.       
In general, as can be seen in Figure~\ref{subfig:noscanspsgluino}) and Figure~\ref{subfig:noscansps3g}, for all SPS$_{NS}$  spectra, as expected, the solid markers lie within our universal exclusion regions (red area), and the open markers lie either in the model-dependent exclusion region (yellow area) or in the allowed region (green area). This confirms that the quoted exclusion limits are also valid for spectra possessing sparticle complexity which are typical of complete SUSY models like the CMSSM.

As a second validation of our results, we now perform a scan in the gluino and third-generation squark mass planes of these SPS$_{NS}$ spectra, keeping all other sparticle masses at their nominal values. As in the case of the NS spectra defined in Section~\ref{sec:spectra}, we derive 95\% exclusion confidence level limits on the masses of the gluino and third-generation squarks. If our limits are indeed universal, and properly take into account uncertainties arising from the underlying complexity, we expect that all of the SPS$_{NS}$ spectra points must lie within the $\pm 1 \sigma$ region (solid black lines) defining the universal limit. As can be seen in Figure~\ref{subfig:noscansps3g}, this is the case for all the SPS$_{NS}$ points, verifying that the uncertainties taken into account when constructing our limits do indeed cover the variety of possible Natural-like SUSY spectra. In particular, the spread of the limits obtained for the SPS$_{NS}$ points is about 100 GeV. This further confirms that our estimate of around 100 GeV for the spectrum dependent systematic uncertainty, as obtained from the systematic variations of NS4 and dominates the total systematic error, is an appropriate uncertainty estimate. Therefore, our limits are applicable to spectra exhibiting a level of complexity similar to those predicted in full SUSY models like the CMSSM. 

These two tests further highlight the fact that our gluino and third-generation squark mass limits are applicable to all Natural-like SUSY models.

\begin{figure*}[htb!]
\centering
\subfigure[]{\resizebox{6cm}{!}{\includegraphics{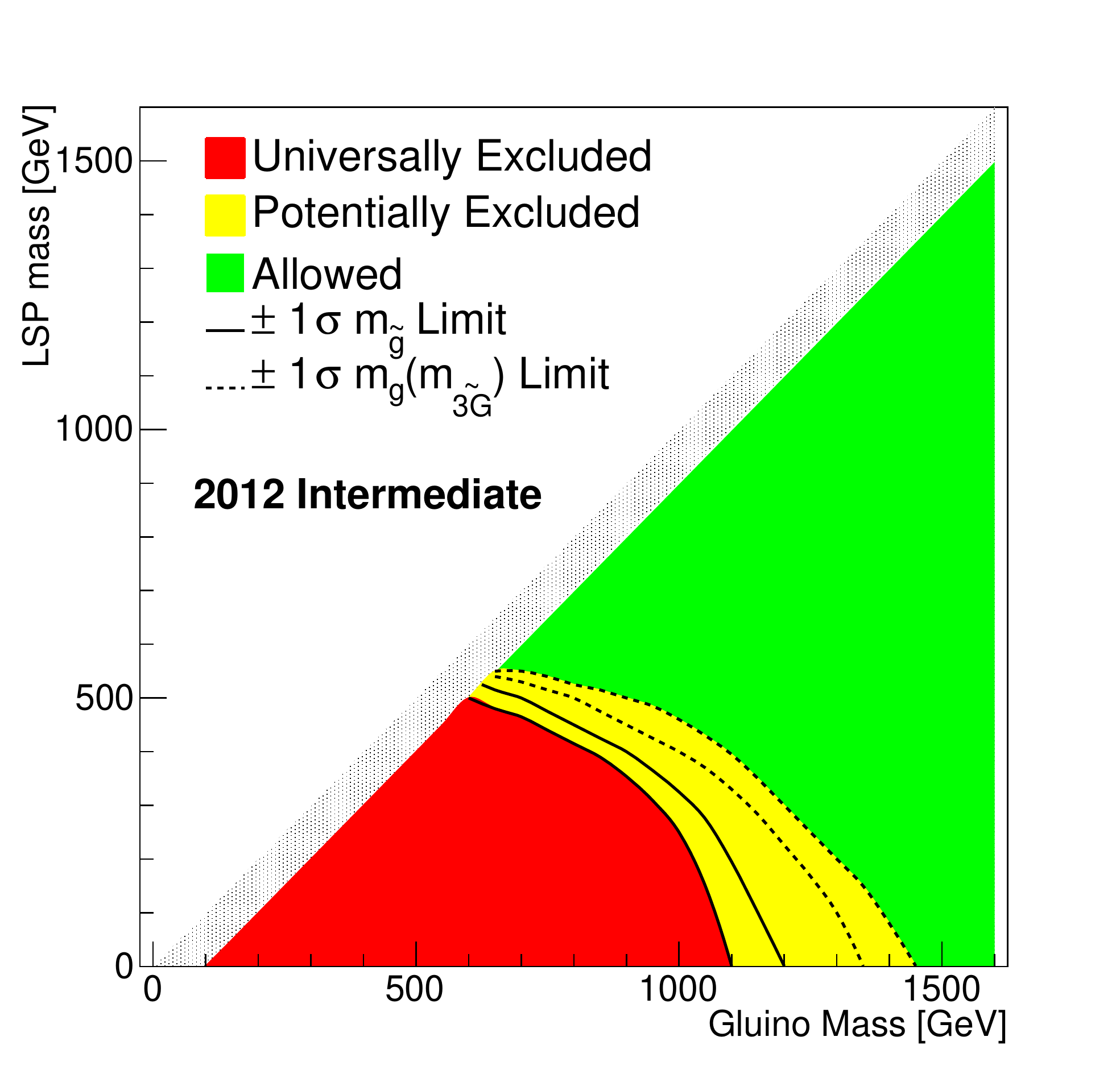}}\label{subfig:mgluino2012i}}
\subfigure[]{\resizebox{6cm}{!}{\includegraphics{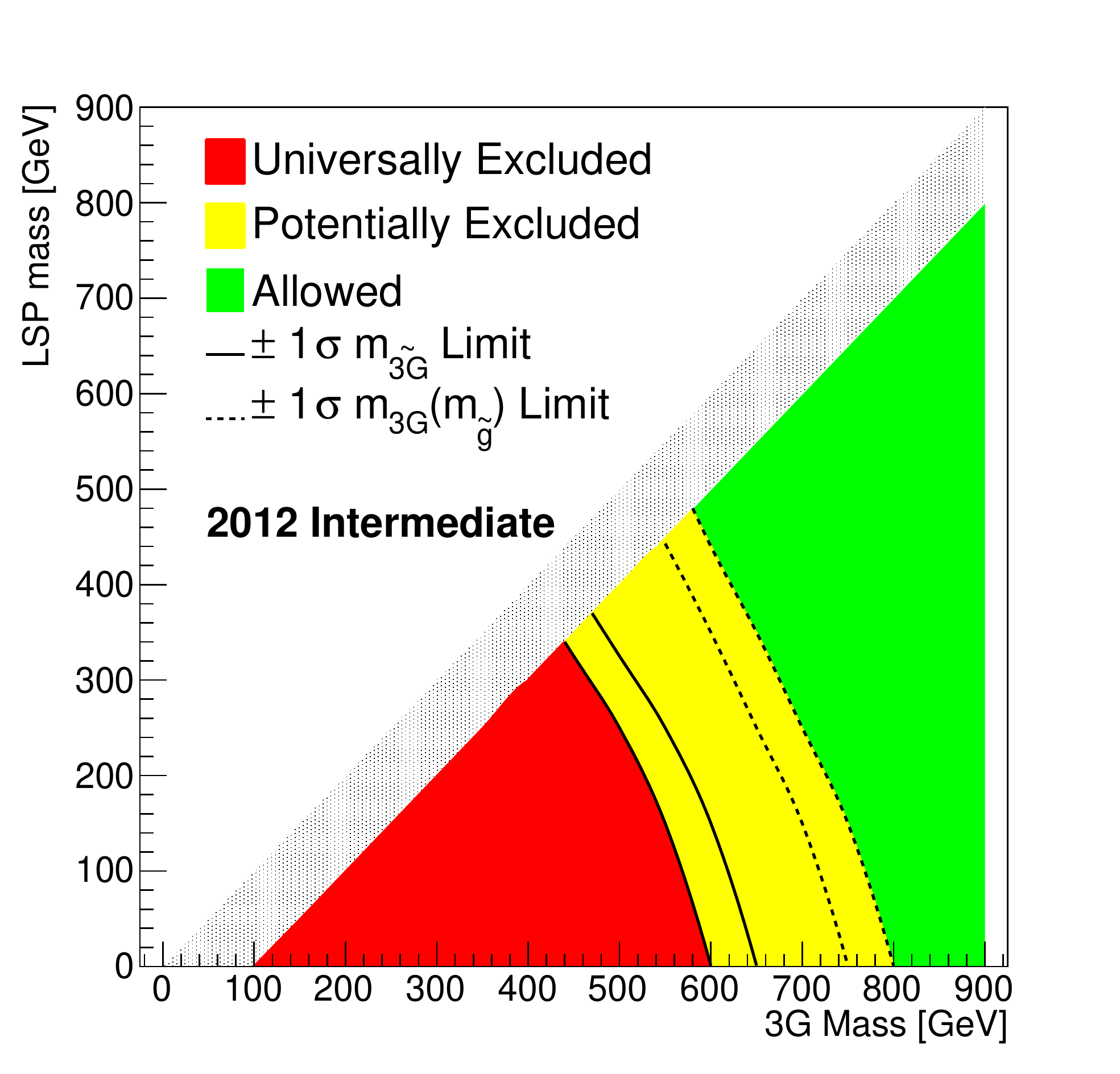}}\label{subfig:m3g2012i}}
\subfigure[]{\resizebox{6cm}{!}{\includegraphics{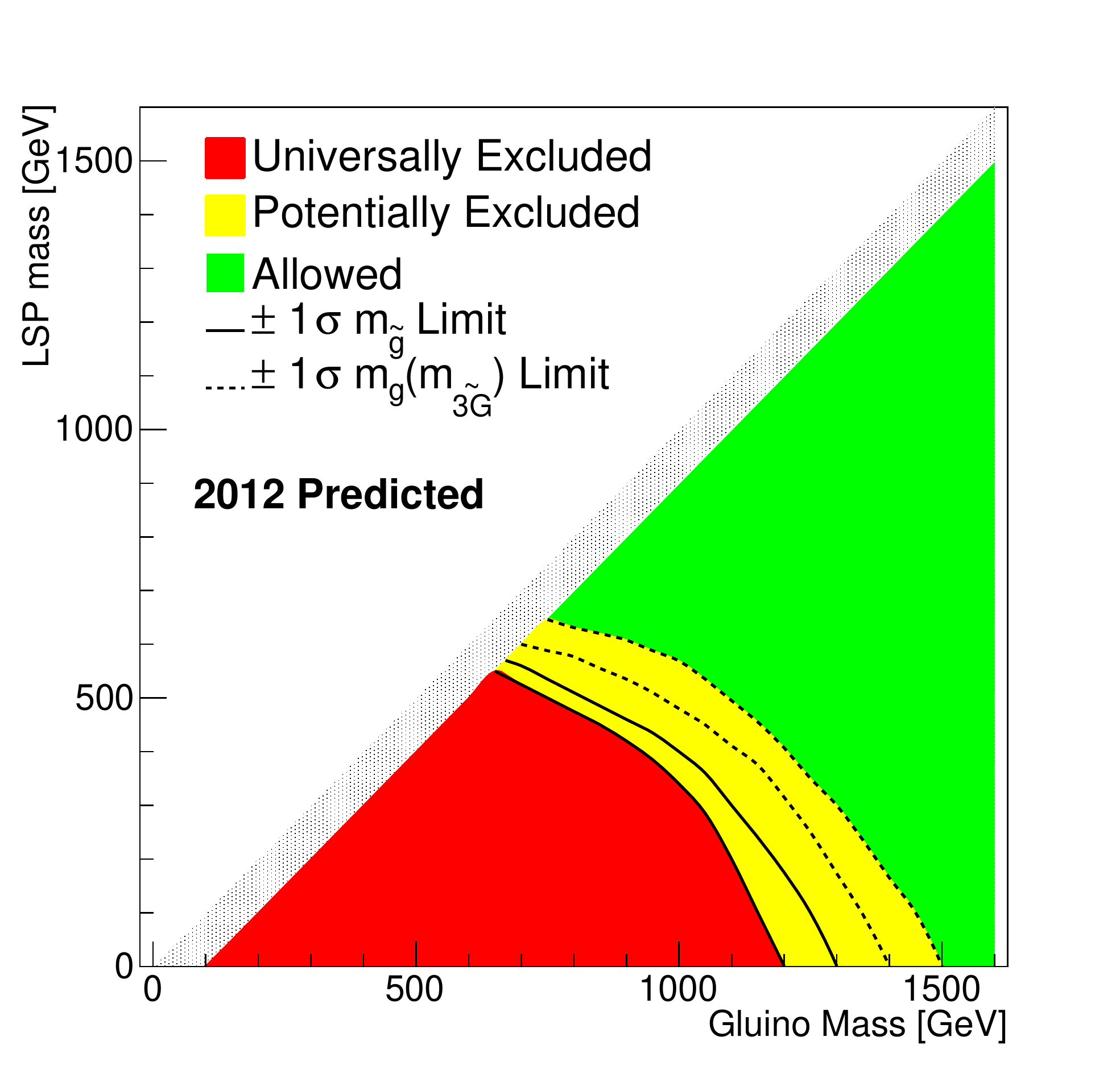}}\label{subfig:mgluino2012}}
\subfigure[]{\resizebox{6cm}{!}{\includegraphics{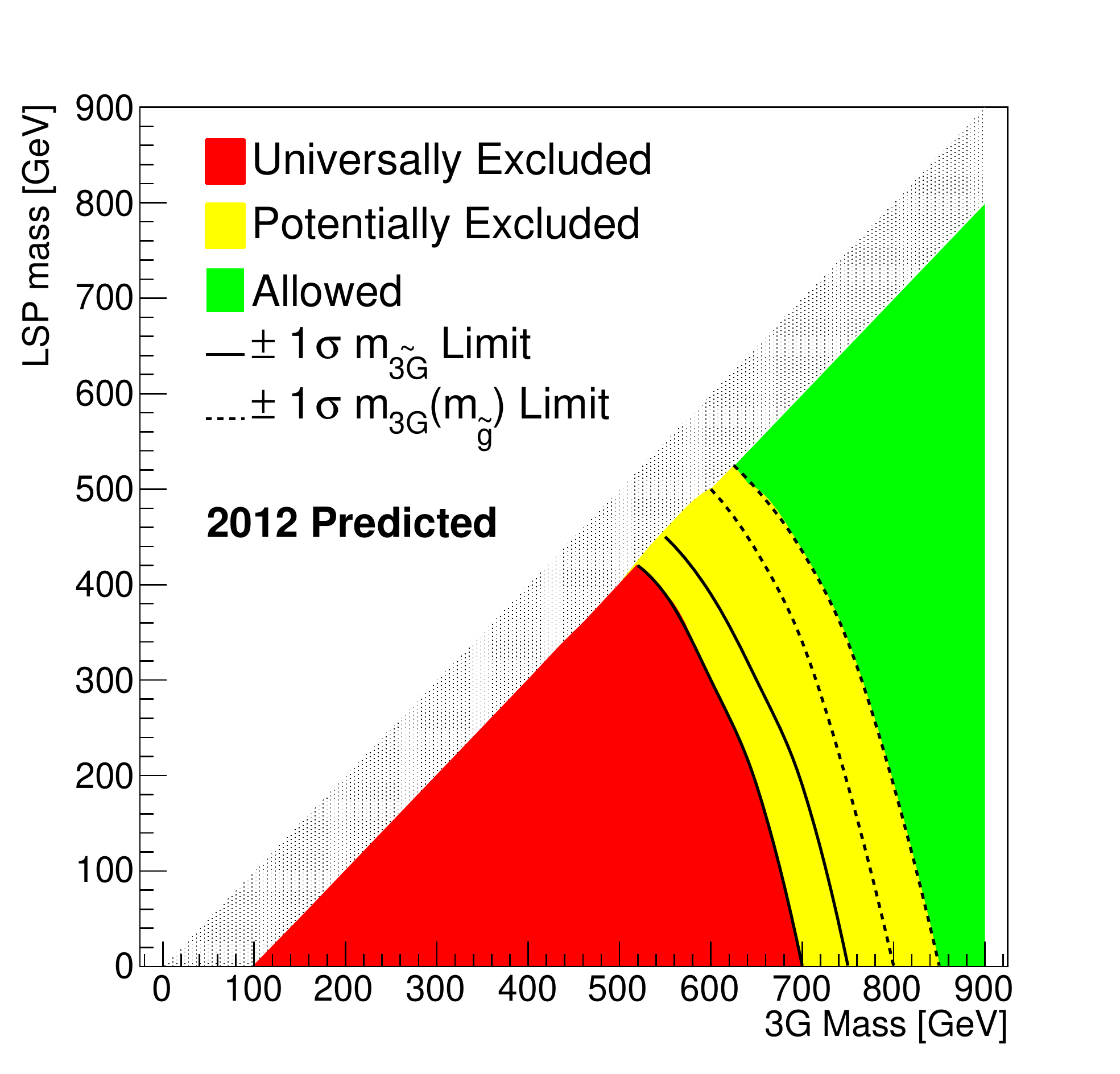}}\label{subfig:m3g2012}}

\caption{\it Gluino and third-generation limit plots for two different 2012 scenarios. Figure~\ref{subfig:mgluino2012i} and  Figure~\ref{subfig:m3g2012i} show the gluino and third-generation mass limits, respectively, for the 2012 intermediate scenario. The corresponding limits for the prediction of the full 2012 dataset, combined between CMS and ATLAS, are shown in Figure~\ref{subfig:mgluino2012} and Figure~\ref{subfig:m3g2012}. For further details on the colour scheme used, see the text in Section~\ref{sec:results2011}.}\label{fig:moneyplot2012}
\end{figure*}

\subsection{Results based on a intermediate set of 2012 searches}\label{sec:results1112}
The analysis as described in Section~\ref{sec:results2011} has been repeated using the available $\sqrt{s}=8$~TeV updates. These correspond to only two out of the four CMS searches considered, the $\alpha_{T}$ search~\cite{Chatrchyan:2013lya} and the same-sign dilepton search~\cite{Chatrchyan:2012paa}, both with approximately $11~\textrm{fb}^{-1}$ of integrated luminosity collected in 2012. These two searches are considered in combination with the 2011 single-lepton and opposite-sign dilepton searches, and the analysis chain is repeated. The results are summarised in the top panels of Figure~\ref{fig:moneyplot2012} and a summary of these exclusion limits is provided in the middle section of Table~\ref{tab:limits}. As can be seen, the addition of the 2012 searches improve the limits by around 100~GeV for low masses of the LSP, and around 50~GeV for high masses of the LSP. 

\subsection{Prediction for the full 2012 data set }
As an outlook for the results of the full analysis from the 2012 dataset, we repeat the exercise in Section~\ref{sec:results1112}, linearly scaling up all searches used in Section~\ref{sec:results1112} to an integrated luminosity of $40~\textrm{fb}^{-1}$ each. Here, we make the assumption based on past experiences, that both the CMS and ATLAS searches possess similar sensitivities. 

The results of our prediction for the ultimate search reach of CMS and ATLAS combined are shown in the bottom panels of Figure~\ref{fig:moneyplot2012}, and a summary of these exclusion limits is provided in the lower third of Table~\ref{tab:limits}. This outlook suggests that for low masses of the LSP, universal limits on gluino and third-generation squarks with masses $m_{\tilde{g}}\approx$ 1150 GeV and $m_{\tilde{3G}}\approx$ 675 GeV are expected. For high LSP masses, these universal limits are expected to decrease to $m_{\tilde{g}}\approx$ 650 GeV and $m_{\tilde{3G}}\approx$ 500 GeV.
  
\subsubsection*{Note added}
While this work was being completed, both the ATLAS and CMS collaborations released preliminary results with SUSY searches based on the entire data set. None of these preliminary results represent an update of any of the chosen inclusive searches for our analysis, and thus at the time of publication, the set of results chosen for this paper are still fully up to date. Despite this, these preliminary results confirm that our predictions for the analysis of the 2012 dataset represents a reasonable approximation to the search reach that can be expected when the CMS and ATLAS results are combined.

\begin{table*}[!tbh!]
\tbl{An overview of the most important limits obtained from our study, shown separately for the different scenarios considered in this paper. The two left-hand columns show the universal limits on gluino and third-generation squark masses at the corresponding $-1 \sigma$ boundary, as defined by the outer perimeter of the red regions in Figure~\ref{fig:spsvalidation} and Figure~\ref{fig:moneyplot2012}. The two right-hand columns feature the spectrum dependent limits defined by the outer perimeter of the yellow area in the same Figures.}{

\begin{tabular}{c|cc||cc| } \cline{2-5}   

&    $m_g$  & $m_{\tilde{3G}}$  & $m_g (m_{\tilde{3G}})$ & $m_{\tilde{3G}} (m_g)$   \\ 
&   for ($m_{LSP}$)  & for ($m_{LSP}$)  & for ($m_{LSP}$) & for ($m_{LSP}$)  \\    \cline{2-5}  \cline{2-5} 
 & \multicolumn{4}{ |c| }{2011 final} \\    \cline{2-5} 
 \multirow{2}{*}{strongest limit}  & $\approx$ 900 GeV & $\approx$  475 GeV & $\approx$ 1250 GeV & $\approx$675 GeV   \\  
 						& ($\leq$ 100 GeV) & ($\leq$  100 GeV) & ($\leq$ 100 GeV) & ($\leq$100 GeV )  \\     \cline{2-5} 
 
 \multirow{2}{*}{weakest limit}  & $\approx$ 550 GeV & $\approx$  400 GeV & $\approx$ 600 GeV & $\approx$550 GeV   \\ 
						& ($\approx$ 450 GeV) & $\approx$  (300 GeV) & ($\approx$ 500 GeV) & ($\approx$450 GeV)   \\   \cline{2-5} \cline{2-5}

 & \multicolumn{4}{ |c| }{2012 intermediate} \\    \cline{2-5} 
 \multirow{2}{*}{strongest limit}  & $\approx$ 1050 GeV & $\approx$  575 GeV & $\approx$ 1400 GeV & $\approx$775 GeV   \\  
 						& ($\leq$ 100 GeV) & ($\leq$  100 GeV) & ($\leq$ 100 GeV) & ($\leq$100 GeV )  \\     \cline{2-5} 
 
 \multirow{2}{*}{weakest limit}  & $\approx$ 600 GeV & $\approx$  450 GeV & $\approx$ 650 GeV & $\approx$600 GeV   \\ 
						& ($\approx$ 500 GeV) & $\approx$  (350 GeV) & ($\approx$ 550 GeV) & ($\approx$500 GeV)   \\   \cline{2-5} \cline{2-5}
						
						& \multicolumn{4}{ |c| }{2012 prediction} \\    \cline{2-5} 
 \multirow{2}{*}{strongest limit}  & $\approx$ 1150 GeV & $\approx$  675 GeV & $\approx$ 1450 GeV & $\approx$800 GeV   \\  
 						& ($\leq$ 100 GeV) & ($\leq$  100 GeV) & ($\leq$ 100 GeV) & ($\leq$100 GeV )  \\     \cline{2-5} 
 
 \multirow{2}{*}{weakest limit}  & $\approx$ 650 GeV & $\approx$  500 GeV & $\approx$ 750 GeV & $\approx$650 GeV   \\ 
						& ($\approx$ 550 GeV) & $\approx$  (400 GeV) & ($\approx$ 650 GeV) & ($\approx$550 GeV)   \\   \cline{2-5} \cline{2-5}

\end{tabular}
\label{tab:limits}}
\end{table*}

\section{Summary and Conclusion}
\label{sec:conclusion}
Using our analysis framework, we have performed a reinterpretation based on a consistent set of four inclusive topology searches from the CMS experiment (zero-lepton, one-lepton, opposite-sign di-lepton, and same-sign di-lepton). Extensive validation studies were carried out to ensure that our framework could reproduce the quoted experimental limits within acceptable uncertainties for each of these searches. We then combined the signal expectations from each of these searches simultaneously to obtain limits on the gluino mass, $m_{\tilde{g}}$, and third-generation squark mass, $m_{\tilde{3G}}$, where the latter represents the numerical average of the considered third-generation squark masses. Our mass limits were calculated using a simple reference spectrum NS4 (see Figure~\ref{subfig:NS4}) which is characteristic of the broad class of Natural-like SUSY spectra for which our limits are valid. In order to account for the important role of the LSP in the signal acceptance of experimental searches, we calculated limits on $m_{\tilde{g}}$ and $m_{\tilde{3G}}$ as a function of the mass of the LSP. 

On the basis of systematic studies for various incarnations of Natural-like SUSY spectra, we have demonstrated that, as expected, the limits of the individual searches exhibit strong dependencies on the underlying spectrum complexity. However, the combination of the inclusive topology searches yield limits that are far more stable with respect to the assumed underlying complexity. In fact, within the uncertainties of our analysis, the results quoted in this paper represent universal limits on the gluino and third-generation squark masses in the context of Natural-like SUSY spectra. To further reinforce this point, we modified several of the historical SPS benchmark points (SPS1a - SPS5) by removing the first and second generation squarks and analysing them through our framework. For each of these points, we find exclusion confidence levels consistent with the expectations from our studies, as dependent on the gluino and third-generation squark masses.

We then provided these universal mass limits in three scenarios, where the first scenario summarises the final results of the 2011 data taking. The second scenario provides a snapshot of the current status, based on a partial set of 2012 searches, and the third scenario represents an outlook of how these limits might evolve with the full 2012 data set with both the CMS and ATLAS experiments combined. To enable a fast interpretation of our limits to an arbitrary Natural-like SUSY spectrum, we define a very simple colour code in our corresponding results, referred to as ``traffic light'' plots, with three distinct regions. In this respect, we provide an alternative approach for interpreting experimental limits, which follow the same format as those limits presented in the SMS scenarios by the LHC experiments. However, in contrast to those SMS limits, our ``traffic light'' plots are not just valid for one simple decay chain, but are instead universal to a broad class of SUSY spectra. Therefore, this approach enables an even broader application of experimental limits without detailed theoretical assumptions. 

For an LSP mass of approximately $100$~GeV, we find that based on the entire $\sqrt{s}=7$~TeV dataset of 2011, the combination of the CMS searches exclude a universal, model independent gluino mass of $m_{g}\approx$ 900 GeV and a third-generation squark mass of $m_{\tilde{3G}}\approx$ 475 GeV. Today, these universal limits are extended to $m_{\tilde{g}}\approx$ 1050 GeV and $m_{\tilde{3G}}\approx$ 575 GeV. For the final result based on the entire 2012 dataset, and with both experiments combined, we expect that these limits could increase to $m_{\tilde{g}}\approx$ 1150~GeV and $m_{\tilde{3G}}\approx$ 675~GeV. 
This suggests that Natural SUSY scenarios based on rather stringent fine-tuning requirements, which predict at least three third-generation squarks with masses well below 1 TeV, a gluino of a mass less than about 1.5 TeV, and higgsinos of masses below 0.5 TeV~\cite{Barbieri198863}, are likely not completely ruled out. Therefore, the search for signals of gluino induced and direct third-generation squark production in the context of Natural-like SUSY, as well as SUSY signatures in general, will continue in 2015 when the LHC will come back online at a higher energy.

Besides quantitatively establishing mass limits on gluino and third-generation squarks in the context of Natural-like SUSY spectra, our work strongly indicates that with the simple yet very powerful ansatz of combining a representative set of inclusive topology SUSY searches, it is possible to determine universal mass limits on coloured sparticles for an even broader class of SUSY spectra. In the case of Natural-like SUSY spectra, we were able to demonstrate that the problem of establishing universal limits based on the combination of topology searches is governed by three parameters, namely the gluino mass ($m_{\tilde{g}}$), the average third-generation squark mass ($m_{\tilde{3G}}$), and the mass of the lightest SUSY particle ($m_{LSP}$). Preliminary studies indicate however, that extending this set of basic parameters to also include the average first and second generation squark mass ($m_{\tilde{12G}}$), will allow us to establish universal limits for an even broader class of SUSY spectra. In fact, it seems possible that interpreting the combination of a consistent set of SUSY searches in the context of these four basic variables ($m_{\tilde{g}}$, $m_{\tilde{12G}}$, $m_{\tilde{3G}}$, $m_{LSP}$) will provide universal limits on coloured sparticles that are valid for almost any arbitrary SUSY spectrum, within reasonable uncertainties. This extension to our work is still in progress and will be subject of another paper.

This conclusion also reinforces the importance of inclusive topology searches in a general SUSY search strategy. These searches play an important role in obtaining both the best, and also the least model dependent limits on sparticle masses. They are also the key ingredient, like they were in 2010 and 2011, for the early phase of LHC running in 2015, where the new energy frontier will once again break new ground. For this reason, we see with some concern the recent increase in emphasis on searches dedicated to very specialised signatures, which often only probe with significance a handful of production and decay modes, e.g. direct or gluino mediated third-generation squark production. While such dedicated searches are useful to complement and refine the general search strategy for SUSY at the LHC, their limited scope and strong model-dependent assumptions should prevent them from becoming the core of the search programme. This role can only be filled by inclusive topology searches, which are defined by key experimental signatures, such as missing transverse energy, multi-jets, jets coming from b-quarks, and possibly isolated leptons and photons in the final state. The return to a SUSY search strategy centred around inclusive topology searches, in which dedicated and specialised searches complement the core of the programme, is crucial. This is especially important as we enter into the preparatory phase for the higher energy running of the LHC in 2015, where we will once again have the best opportunity for discovery!

\section*{Acknowledgements}
The authors would like to thank members of the MasterCode collaboration for very useful discussions. The work of the authors is supported in part by the London Centre for Terauniverse Studies (LCTS), using funding from the European Research Council via the Advanced Investigator Grant 267352. 

\bibliography{nspaper_OBJM3_v1}{}
\bibliographystyle{ws-ijmpa}
\end{document}